\documentstyle[aps]{revtex}
\begin{document}
%\draft
\newtheorem{ex}{Example}
\newtheorem{lem}{Lemma}

\title{
Elements of Nonlinear Quantum Mechanics (I):\\
Nonlinear Schr\"odinger equation and two-level atoms
}
\author{Marek Czachor~\cite{*}}
\address{
Pracownia Dielektryk\'{o}w i P\'{o}\l przewodnik\'{o}w
Organicznych\\
Wydzia{\l}  Fizyki Technicznej i Matematyki Stosowanej\\
 Politechnika Gda\'{n}ska\\
ul. Narutowicza 11/12, 08-952 Gda\'{n}sk, Poland
}

\maketitle
\begin{abstract}
Starting with the same form of atomic nonlinearity Weinberg
\cite{W2} and W\'odkiewicz and Scully \cite{WScully}
obtained contradictory results concerning an evolution of the
atomic inversion $w$ in a two-level atom in Weinberg's nonlinear
quantum mechanics: If the atom is initially in a ground state
then either (1) the evolution of $w$ can be linear if one uses a
nonlinear generalization of the
Jaynes-Cummings Hamiltonian, or (2) is always nonlinear if
one uses the nonlinear Bloch
equations derived from the nonlinear atomic Hamiltonian
function.
It is shown that the difference is rooted in inequivalent
descriptions of the composite ``atom+field" system. The linear
evolution of $w$ results from a ``faster-than-light
communication" between the atom and the field. If one applies a
description without the ``faster-than-light telegraph" then the
calculations based on a suitably modified Jaynes-Cummings
Hamiltonian lead to the same dynamics of $w$ as this found in
semiclassical calculations based on Bloch equations.
A physical background of the two approaches is discussed in
detail, and two, inequivalent kinds of such ``telegraphs" are described.
 It is shown  also that a nonlinear quantum mechanics
based on a nonlinear Schr\"odinger equation does not possess a
natural probability interpretation. Various definitions of
nonlinear eigenvalues and probabilities are discussed and
illustrated with examples.
Finally, I discuss problems that appear if we apply the above
results to the ``object+observer" composite system, where the
``observer" can evolve in a nonlinear way.
\end{abstract}
\section{Introduction}

The purpose of the generalization of quantum mechanics proposed
by S.~Weinberg
\cite{W2,W,W1} ``was not
seriously to propose an alternative to quantum mechanics but only to
have {\em some\/} theory whose predictions would be close to but not
quite the same as those of quantum mechanics, to serve as a foil that
might be tested experimentally" \cite{Dreams}. Next, in a series of
very precise  experiments the Weinberg's theory was indeed tested
\cite{Exp}, showing no observable deviations from  linearity.
 This was neither
particularly surprising, as Weinberg admits in his recent
popular book, nor the most discouraging feature of
the new framework: ``Even if there are small
nonlinear corrections to quantum mechanics, there was no reason to
believe that these corrections should be just large enough to show up
in the first round of experiments designed to search for them. What I
did find disappointing was that this nonlinear alternative to quantum
mechanics turned out to have purely theoretical internal difficulties.
For one thing, I could not find any way to extend the nonlinear version
of quantum mechanics to theories based on Einstein's special theory of
relativity. Then, after my work was published, both N.~Gisin in Geneva
and my colleague Joseph Polchinski at the University of Texas
independently pointed out that in the Einstein-Podolsky-Rosen thought
experiment (...) the nonlinearities of the generalized theory {\em
could\/} be used to send signals instantaneously over large distances,
a result forbidden by special relativity. At least for the present I
have given up the problem; I simply do not know how to change quantum
mechanics by a small amount without wrecking it altogether"\cite{Dreams}.

This quotation  explains
almost all purposes of this paper. In its first part I will try
to point out the main {\em internal theoretical difficulties\/}
of those versions of nonlinear quantum mechanics (NLQM)
that are based on a
nonlinear Schr\"odinger equation.
We shall see, in particular, that the problem of a correct description of
composite systems in NLQM is not only related to the celebrated
``faster-than-light telegraphs" mentioned in the above
quotation, but may also lead to different predictions for
 two-level atoms, a basic theoretical tool used in calculations
for actual experiments.

Just to illustrate the latter  let us notice the following
apparent inconsistency. It has been shown in \cite{W2} that
 a system consisting initially of a
single photon electromagnetic field and a two-level atom in a
ground state, performs ordinary {\em linear\/} Rabi oscillations
even in case the atomic Schr\"odinger equation contains a
nonlinear term. On the other hand, the calcultions
of W\'odkiewicz and Scully  \cite{WScully}
 starting with
the same form of nonlinearity but based on the Bloch equations
approach
lead, with the same initial condition for the atomic
inversion, to elliptic hence {\em nonlinear\/} oscillations of
the inversion. It
will be shown below that the difference is rooted simply in
non-equivalent descriptions of the composite ``atom+field"
system. The linear evolution occurs if we use the description
with an implicit ``faster-than-light telegraph", whereas the
nonlinear one is found if no such ``telegraphs" are
present.

The two descriptions are based on different physical
assumptions which shall be discussed in detail below.
To make matters worse, it will be shown  that there does not
exist a simple alternative between the two ways of describing
compound systems. In fact, we will see that there exists an
{\em infinite\/} number of {\em inequivalent\/} and still
``correct" possibilities of describing systems such as the
the atom and the field, if one assumes that the atom alone evolves
according to some nonlinear Schr\"odinger equation. This
arbitrariness not only leads to inequivalent descriptions of
optical phenomena but also to fundamental difficulties in
interpretation of experiments.

Another non-unique element which
additionally obscures the meaning of experiments is related to
the probabilistic interepretation of NLQM based on a nonlinear
Schr\"odinger equation. I will show that there does not exist a
unique definition of eigenvalues or probabilities
(propositions), and those used for interpreting experiments testing
Weinberg's theory were not applied in an internally consistent
way.

In the second part of the paper \cite{II} I will propose a new framework
for NLQM. The proposed generalization is based on  a kind of
Nambu mechanics where nonlinearities can be introduced without
any modification of observables. This formulation of NLQM will
be free from the difficulties of the previous approaches, no
``faster-than-light telegraphs" will occur, the description of
composite systems and the probalistic interpretation will be
unique, and, last but not least,
 such a theory can be naturally formulated
in a relativistic way. The results of both parts are based on
the dissertation \cite{doktorat}.

The following sections are devoted to a critical presentation of
the basic theoretical concepts of NLQM.

\section{Nonlinear Quantum Mechanics {\em \'a la} Kibble,
Weinberg, Polchinski, and Jordan}
\label{WEINBERG}

Adding a nonlinear term to the Schr\"odinger equation does not
mean that we have a generalization of quantum mechanics. An
evolution equation is only a part of a theory. A
complete theory must give a concrete prescription for
a translation of theoretical formulas into the language of
experiment.
In the context of quantum theories we must know first of all
what are the mathematical cunterparts of random variables,
probabilities, and averages measured in a laboratory. In linear quantum
mechanics (QM) they are represented by eigenvalues and matrix
elements of projectors or self-adjoint operators. In NLQM based
on nonlinear Schr\"odinger equation none of those objects
appears in a natural way.

\subsection{States}

We will discuss theories where states are represented in the
same way as in ordinary quantum mechanics (the so-called normal
states): by vectors or rays in a Hilbert space, or by density
matrices.  We should also keep in mind that possible extensions
of QM may involve different manifolds of states \cite{Kibble,K2},
such as general K\"ahler manifolds
\cite{CMP1,CMP2,CMP3,CMP4,CMP5,CMP6}.  A
necessity of some extension of that kind is suggested, for
example, by Penrose
\cite{Emperror}. We will not include stochastic
generalizations either, cf.~\cite{Gisin-st}.

\subsection{Observables}\label{Observables}

In Hamiltonian formulation of the Schr\"dinger equation
(real and imaginary parts of)
components of a state vector play a role of symplectic
coordinates, and observables are bilinear functionals $\langle
\psi| \hat A|\psi\rangle$, where $\hat A$ is a self-adjoint
operator.

The fact that quantum observables are so specifically chosen
leads to a special, complex Hilbertian type of the quantum
mechanical probablity calculus. It seems it was Mielnik who
noticed
\cite{M,M1,M2,MielnikMob} that since there
 exist many different and inequivalent probability models
they may be related to some nonlinear version of quantum
maechanics.
Such a perspective seems quite natural and suggests that one
should try to investigate theories where obsevables, like in
classical mechanics, belong to a more general set of
functionals.

In the generalization proposed by Kibble\cite{Kibble,K2}
 the set of observables consists of
differentiable functions defined on a projective space. In the
Weinberg formulation \cite{W2} one assumes that the observables
are defined on a Hilbert space but satisfy the (1,1)-homogeneity
condition
\begin{equation}
A(\lambda \psi,\bar \psi)=A( \psi, \lambda\bar \psi)=\lambda
A(\psi,\bar \psi).\label{homo}
\end{equation}
Averages are defined in \cite{W2} like in ordinary quantum
mechanics. The (1,1)-homogeneity and the definition of averages
introduces the projective structure into the set of states and,
effectively, into the dynamics. This is the way the Weinberg
approach can be regarded as a variant of the Kibble's one.  (We
shall see later that Weinberg's theory contains also some
elements that are new with respect to the Kibble's proposal.)

The homogeneity condition (\ref{homo}) can be expressed by means
of the Euler formula as follows
\begin{equation}
\psi_n\frac{\partial A}{\partial\psi_n}=\bar \psi_n\frac{\partial
A}{\partial\bar \psi_n}= A.\label{Euler}
\end{equation}
With the help of (\ref{Euler}) we can deduce some general
properties of the observables. Applying (\ref{Euler}) twice we
find that
\begin{equation}
A=\frac{\partial A}{\partial\psi_n}\psi_n=
\bar \psi_m\frac{\partial^2 A}
{\partial\bar \psi_m\partial\psi_n}\psi_n=
\bar \psi_m\hat A_{mn}\psi_n=\langle\psi|\hat A(\psi,\bar
\psi)|\psi\rangle\label{nl macierz}
\end{equation}
and
\begin{equation}
\frac{\partial A}{\partial\bar \psi_m}=
\frac{\partial^2 A}{\partial\bar \psi_m\partial\psi_n}\psi_n=
(\hat A(\psi,\bar \psi)|\psi\rangle)_m.\label{nl macierz na
wektor}
\end{equation}
We can see that the homogeneity condition leads naturally to a
nonlinear operator algebra. A matrix multiplication of such
nonlinear operators can be easily expressed in terms of first
partial derivatives of observables since
\begin{equation}
A*B:=
\frac{\partial A}{\partial \psi_m}\frac{\partial B}
{\partial\bar \psi_m}=\langle\psi|\hat A(\psi,\bar \psi)\hat B
(\psi,\bar \psi)|\psi\rangle.\label{star}
\end{equation}
This $*$-product conserves the (1,1)-homogeneity of observables
because the nonlinear matrices defined in (\ref{nl macierz}) are
(0,0)-homo\-geneous. The (0,0)-homo\-geneity implies, on the
other hand, that the matrices are, in certain sense, ``almost
constant". Indeed,
\begin{equation}
\Bigl(\frac{\partial\hat A}{\partial\psi_k}|\psi\rangle\Bigr)_l=
\frac{\partial}{\partial\psi_k}\Bigl(
\frac{\partial^2 A}{\partial\bar \psi_l\partial\psi_m}\Bigr)
\psi_m=
\psi_m\frac{\partial}{\partial\psi_m}\frac{\partial^2 A}
{\partial\bar
\psi_l\partial\psi_k}= 0\label{"staly"}
\end{equation}
by virtue of the 0-homo\-geneity Euler condition. Consider now a
differentiable function $t\to\psi_n(t)$. Eq. (\ref{"staly"}) and
its complex conjugate imply that
$
\langle\psi|\frac{d}{dt}\hat A(\psi,\bar \psi)|\psi\rangle=0
$
for any observable $A$. The algebra of observables equippedOB with
the above $*$-product possesses a natural left and right unit
element $n(\psi,\bar \psi)={\langle\psi|\psi\rangle}$.

Despite some similarities to ordinary linear operator algebra
our algebra of observables is associative if and only if the
operators are linear. This leads to several differences with
respect to ordinary quantum mechanics. For example, $*$-powers
of $A$ will not, in general, $*$-commute with $A$, a fact that
will influence the integrability of the theory.

In linear QM the $*$-powers of an observable are related to
higher moments of random variables measured in experiments,
hence lead to the probabilistic interpretation of the theory. In
nonlinear theory the product (\ref{star}) leads to the first
non-unique element of the formalism we shall meet later.

For let us consider an observable $A$. We know that it can be
written in a form of a matrix element $\langle\psi| \hat
A|\psi\rangle$ where $\hat A$ is the nonlinear operator defined
in (\ref{nl macierz}) (for simplicity we shall, from now on,
omit in our notation the dependence of $\hat A$ on $\psi$). The
$*$-square of $A$ can be defined as
$\frac{\partial A}{\partial \psi_m}\frac{\partial A}
{\partial\bar \psi_m}$ or $\langle\psi|\hat A\hat
A|\psi\rangle$.  If we want to
define the third power we encounter several possibilities.  The
apparently most natural choice of $(A*A)*A$ or $A*(A*A)$ is not
acceptable as they are in general complex and non equal
even if $A$ is real.
 In Subsec.
\ref{eigen} we shall see also that $*$ does not lead to a
consistent probability interpretation of $A$. The second
definition can be chosen as $\langle\psi|\hat A\hat A\hat
A|\psi\rangle:=A\bar *A \bar *A$. We see that once we have
defined $\hat A$, which is unique, all its $\bar *$-powers are
unique as well.  Notice that since $A*B=A\bar * B$ both products
lead to the same Poisson bracket  (generate the same
evolution of observables).

One of the most important questions of nonlinear quantum
mechanics concerns a definition of observables that correspond
to subsystems of a larger system. Let us begin the discussion
with recalling the ways we do it in ordinary QM.

 By saying
that $H(\psi_{\rm sub},\bar \psi_{\rm sub})$ is a
Hamilitonian function of a subsystem we mean that $H(\psi_{\rm
sub},\bar
\psi_{\rm sub})$ describes a subsystem which is
noninteracting and noncorrelated with the rest of the Universe
(otherwise the subsystem is not described by a state vector).
Quantum mechanically this means that the state of the Universe
is $|\psi\rangle=| {\psi_{\rm sub}}\rangle \otimes|
{\psi_{\rm rest}}\rangle $ (or $\rho=|
{\psi_{\rm sub}}\rangle \langle
{\psi_{\rm sub}}|\otimes
\rho_{rest}$).

Subsystem observables in linear QM can be written as
\begin{equation}
A(\psi,\bar \psi)=
\langle\psi| \hat A\otimes {\bf 1}_{\rm rest}|
\psi\rangle={\rm Tr} \rho_{\rm
sub}\hat A.
\end{equation}
Let now $\{| r\rangle \}$ denote a basis in the Hilbert space of the
``rest" and $\{| a\rangle \}$ be the one of the subsystem. A general
form of the state of the whole system (here we assume for
inessential simplicity that the whole system is in a pure state)
is
\begin{equation}
|\psi\rangle=\sum_{a,r}\psi_{a,r}| a\rangle | r\rangle =\sum_r|
{\Phi^{(r)}}\rangle | r\rangle .\label{whole st}
\end{equation}
An average of $A$ in the state $|\psi\rangle$ can be expressed
by means of the decomposition (\ref{whole st}) as follows
\begin{eqnarray}
{\overline{A}}(\psi,\bar \psi)&=&\frac{
\langle\psi| \hat A\otimes {\bf 1}_{\rm
rest}|\psi\rangle}{{\langle\psi|\psi\rangle}}=
\sum_r \frac{\langle {\Phi^{(r)}}|\hat A| {\Phi^{(r)}}\rangle }
{\langle {\Phi^{(r)}}| {\Phi^{(r)}}\rangle}
\frac{\langle {\Phi^{(r)}}|
{\Phi^{(r)}}\rangle}{{\langle\psi|\psi\rangle}}\nonumber\\
&=&
\sum_r \frac{\langle {\Phi^{(r)}}|\hat A| {\Phi^{(r)}}\rangle }
{\langle {\Phi^{(r)}}| {\Phi^{(r)}}\rangle}
\frac{\langle\psi|\bigl( {\bf 1}_{\rm sub}\otimes |r\rangle\langle
r|\bigr)|\psi\rangle}{{\langle\psi|\psi\rangle}}.\label{r-sub}
\end{eqnarray}
The physical meaning of (\ref{r-sub}) is obvious: The whole
ensemble of subsystems is decomposed into subensembles
corresponding to different results of measurements of an
observable whose spectral family is given by $\{| r\rangle
\langle r|\}$,
and the average of $A$ in the subsystem is calculated for each
subensemble separately. The form (\ref{r-sub}) is useful as an
illustration of the ``collapse" of a state vector phenomenon.
Indeed, the average is calculated as though the ensemble of
subsystems consisted of subensembles collapsed by an external
observer and, in certain sense, the ensemble is treated as the
one composed of individual objects each of them posessing some
``property" measured ({\it via\/} correlations) by the external
observer.

The description in terms of the reduced density matrix
$\rho_{\rm sub}$ suggests a slightly different interpretation
of the average. The form
\begin{equation}
{\overline{A}}(\psi,\bar \psi)={\rm Tr}\rho_{\rm sub}\hat
A\label{rho-sub}
\end{equation}
means that we do not take into account possible decompositions
of the ensemble, but treat it as a whole.  The presence of the
reduced density matrix in (\ref{rho-sub}) does not mean, of
course, that we cannot use the state vector's components --- it
simply restricts the form of observables to only some functions
of $\psi_{a,r}$.

Quantum mechanics when viewed from the first perspective looks
as a theory of individual systems but with some form of
fundamental limitation of knowledge about states of such
individuals. The second possibility turns QM into a kind of a
``mean field theory". Still, since both forms of description are
physically indistinguishable such discussions whithin the linear
framework seem purely academic.

In nonlinear theories the situation is drastically changed. The
nonlinearity is supposed to be fundamental, hence applying to
each individual even if the individual is well isolated from
other physical systems and no mean-field approximation is
justified.

The problem is practically the following. Assuming that we know
a form of an observable describing a given subsystem itself
(that is, if no interactions or correlations with the ``rest"
exist), what is the form of the observable if there {\it do\/}
exist some interactions or correlations with the ``rest"?

The first, explicitly non-mean-field-theoretic, guess for the
subsystem's (additive) observable is suggested by the
``collapse-like" form (\ref{r-sub}) of the average, i.e.
\begin{equation}
A_{\rm sub}(\psi,\bar \psi)=
\sum_r A(\Phi^{(r)},\bar \Phi^{(r)})\label{W-sub}
\end{equation}
where $A$ is a given function defined on a Hilbert space of the
subsystem and the other definitions are like in (\ref{whole
st}). It is this definition that was chosen by Weinberg in his
formulation of nonlinear QM. For non-bilinear $A(\Phi^{(r)},\bar
\Phi^{(r)})$ the {\it form\/} and {\it value\/} of (\ref{W-sub})
depend on the choice of the basis $\{ | r\rangle \}$.

\begin{ex}\em  Consider the following non-bilinear observable
\begin{equation}
A(\psi,\bar \psi) =\frac{\langle \psi|\sigma_3| \psi\rangle
^2}{\langle
\psi|\psi\rangle}=
\frac{(|\psi_1|^2
-|\psi_2|^2)^2}{|\psi_1|^2+|\psi_2|^2}\label{kanon}
\end{equation}
where $\sigma_3$ is the Pauli matrix and $|\psi\rangle$ belongs to
a two-dimensional Hilbert space. Let the whole system be
described by $|\psi\rangle=\sum_{a=1}^{2}\sum_r\psi_{a,r}|
a\rangle_1| r\rangle_2$. Definition (\ref{W-sub}) yields
\begin{equation}
A(\psi_{\rm sub},\bar \psi_{\rm sub})=
\sum_r
\frac{(|\psi_{1r}|^2
-|\psi_{2r}|^2)^2}{|\psi_{1r}|^2+|\psi_{2r}|^2}
\end{equation}
Let
\begin{equation}
|\psi\rangle=
\underbrace{\psi_{1,1}| 1\rangle_1}_{| {\Phi^{(1)}}\rangle}| 1\rangle_2+
\underbrace{\psi_{2,2}
| 2\rangle_1}_{| {\Phi^{(2)}}\rangle}| 2\rangle_2.
\end{equation}
In this particular state
\begin{equation}
A(\psi_{\rm sub},\bar \psi_{\rm
sub})=|\psi_{11}|^2+|\psi_{22}|^2
\end{equation}
so that each ``collapsed" subensemble contributes to the
observable separately. Let us choose now a different basis
$\{| {r'}\rangle_2\}$ in the Hilbert space of the ``rest", such that
$| 1\rangle_2 = {1\over \sqrt 2} (| {1'}\rangle_2 + |
{2'}\rangle_2)$, $|
2\rangle_2 = {1\over \sqrt 2} (| {1'}\rangle_2 - | {2'}\rangle_2)$. Then
\begin{equation}
|\psi\rangle=
\underbrace{{1\over \sqrt 2}\bigl(\psi_{11}|
1\rangle_1+\psi_{22}| 2\rangle_1\bigr)}
_{| {\Phi^{(1)'}}\rangle}| {1'}|_2+
\underbrace{{1\over \sqrt 2}\bigl(\psi_{11}|
1\rangle_1-\psi_{22}| 2\rangle_1\bigr)}_
{| {\Phi^{(2)'}}\rangle}| {2'}\rangle_2.
\end{equation}
Substitution of $| {\Phi^{(1)'}}\rangle$ and $| {\Phi^{(2)'}}\rangle$
into (\ref{W-sub}) gives
\begin{eqnarray}
A(\psi_{\rm sub},\bar \psi_{\rm sub})&=& {1\over
2}\frac{(|\psi_{11}|^2
-|\psi_{22}|^2)^2}{|\psi_{11}|^2+|\psi_{22}|^2}+ {1\over
2}\frac{(|\psi_{11}|^2
-|\psi_{22}|^2)^2}{|\psi_{11}|^2+|\psi_{22}|^2}\nonumber \\
&=&\frac{(|\psi_{11}|^2
-|\psi_{22}|^2)^2}{|\psi_{11}|^2+|\psi_{22}|^2}
\end{eqnarray}
which is also a sum of contributions from two subensembles, but
now ``collapsed to linear polarizations".$\Box$\end{ex}

The above example reveals two properties of (\ref{W-sub}):
\begin{enumerate}
\item This form is basis-dependedent so one should rather write
${A_{\rm sub}}(\psi,\bar \psi) =A_{\rm sub}(\psi,\bar
\psi,\{| r\rangle\})$;

\item If the ``rest" is described by linear QM then no change of
$\{| r\rangle\}$ can influece observables that correspond to ``the
rest" itself. In particular, if one considers a total energy of
the whole system then a change of $\{| r\rangle\}$ will influence
its value (we change the energy of the subsystem and maintain
the energy of the ``rest"). It follows that the form
(\ref{W-sub}) can apply only to systems that, as a whole, are
open (that is, there must exist also some another {\it
nonlinear\/} system not contained in the ``whole" one and
compensating the changes of energy due to the changes of $\{|
r\rangle\}$).

\item The ``subsystem+compensating subsystem+rest" system must
be described in a basis-independent way.
\end{enumerate}

The last remark indicates that in order to describe systems with
``isolated" nonlinearities, that is with the ``rest" linear, one
has to describe subsystems in a way different from Weinberg's.
The following modification was proposed by Polchinski
\cite{Polchinski}.

Consider a nonlinear observable $A(\psi,\bar \psi)$ corresponding to some
isolated physical system, and assume that the observable can be
expressed as a function of the density matrix $\rho=|
\psi\rangle\langle \psi|$, i.e. $A(\psi,\bar \psi)=A(\rho)$
(this is always true in linear QM). For
example, the
observable from the previous example could be written in either
of the forms
\begin{equation}
\frac{({\rm Tr}\,\rho \sigma_3)^2}{{\rm Tr}\,\rho},
\end{equation}
\begin{equation}
\frac{{\rm Tr}\,(\rho\sigma_3\rho\sigma_3)}{{\rm Tr}\,\rho},
\end{equation}
\begin{equation}
\frac{({\rm Tr}\,\rho\sigma_3)^2
{\rm Tr}\,(\rho^2)}{({\rm Tr}\,\rho)^3},
\end{equation}
or
\begin{equation}
\frac{({\rm Tr}\,(\rho\sigma_3\rho\sigma_3))^{n+1}}{{\rm
Tr}\,\rho ({\rm Tr}\,\rho
\sigma_3)^{2n}},
\end{equation}
and so on. For $\rho=|\psi\rangle\langle \psi|$ all of them, and all
their convex combinations, reduce to (\ref{kanon}).
The map $\rho \to A(\rho)$ determines $A(\psi,\bar \psi)$
uniquely, but from the knowledge of $A(\psi,\bar \psi)$ one
cannot deduce the form of $A(\rho)$ for $\rho$ different from
$|\psi\rangle\langle \psi|$. This trivial remark has important
consequences. Returning to the ``atom+field" example,
we see that the knowledge of the nonlinear Schr\"odinger
equation for the
atom alone does not determine uniquely the form of the evolution
equation of the ``atom+field" composite system! The nonlinear
equation describing the atom {\em if there are no correlations
with the field\/} is determined by the Hamiltonian function
$H_{AT}(\psi,\bar \psi)$, but this function no longer describes
the atom if some correlations with the field do appear.

Therefore, it becomes clear that, in order to have a unique
formalism, one should start from the very beginning with the
description of observables in terms of density matrices. Such a
reformulation of Weinbrg's theory was proposed by Polchinski and
Jordan~\cite{Polchinski,Jordan}. In the second part of this
paper I will  generalize the density matrix
formalism, so here we shall concentrate only on its
interpretational problems.

\begin{ex}\em  Let $|\psi\rangle=\psi_{11}| 1\rangle_1| 1\rangle_2+
\psi_{22}| 2\rangle_1| 2\rangle_2$. The reduced density matrix
representing the subsystem is
\begin{equation}
\rho=|\psi_{11}|^2 | 1\rangle_1\langle 1|_1+
|\psi_{22}|^2 | 2\rangle_1\langle 2|_1.
\end{equation}
The subsystem's observables are now
\begin{equation}
\frac{({\rm Tr}\,\rho \sigma_3)^2}{{\rm Tr}\,\rho}=
\frac{(|\psi_{11}|^2
-|\psi_{22}|^2)^2}{|\psi_{11}|^2+|\psi_{22}|^2},\label{meanlike}
\end{equation}
\begin{equation}
\frac{{\rm Tr}\,(\rho\sigma_3\rho\sigma_3)}{{\rm Tr}\,\rho}
=\frac{|\psi_{11}|^4
+|\psi_{22}|^4}{|\psi_{11}|^2+|\psi_{22}|^2},\label{collapselike}
\end{equation}
\begin{equation}
\frac{({\rm Tr}\,\rho\sigma_3)^2{\rm Tr}\,(\rho^2)}
{({\rm Tr}\,\rho)^3}=
\frac{(|\psi_{11}|^2
-|\psi_{22}|^2)^2(|\psi_{11}|^4 +|\psi_{22}|^4)}
{(|\psi_{11}|^2+|\psi_{22}|^2)^3},
\end{equation}
and
\begin{equation}
\frac{({\rm Tr}\,(\rho\sigma_3\rho\sigma_3))^{n+1}}{{\rm
Tr}\,\rho ({\rm Tr}\,\rho
\sigma_3)^{2n}}=
\frac{(|\psi_{11}|^4
+|\psi_{22}|^4)^{n+1}}
{(|\psi_{11}|^2+|\psi_{22}|^2)(|\psi_{11}|^2-|
\psi_{22}|^2)^{2n}}.
\label{singularlike}
\end{equation}
As we can see the observables are completely different. If
$|\psi_{11}|=|\psi_{22}|$ (as in the spin-1/2 singlet state)
the average of the first and third observables vanishes, is
non-vanishing but finite in the second case, and infinite in the
last one.
\end{ex}

Expression (\ref{meanlike}) seems, at first glance, the most
natural one since can be rewritten as
\begin{equation}
\frac{\langle\psi|\sigma_3\otimes {\bf
1}_{rest}|\psi\rangle^2}{{\langle\psi|\psi\rangle}}.
\end{equation}
Kibble in \cite{Kibble} did not explicitly define his way of
describing subsystems, but the examples he discussed suggest
this type of description. In the next section we will discuss
problems with nonlocality of nonlinear QM and see that all the
descriptions by means of reduced density matrices are more safe
than the proposal of Weinberg. However, if some correlations
between subsystems actually exist it seems we indeed {\it can\/}
``collapse" a state of a remote individual. We may expect that
an average value of this ``remote" observable should be composed
of the discussed averages of subensembles. The description by
means of (\ref{meanlike}) does not seem to possess this
property: It is the ``average spin" that contributes to the
observable (\ref{meanlike}). So we first calculate the
``average" of $\sigma_3$ and then square it. In the Weinberg
description we first square the ``spin" and then take the
``average". (The quotation marks are necessary because we have
not defined yet what we mean by eigenvalues, or values of single
measurements; there exists also another subtle point that has
been ignored in the definition of (\ref{W-sub}) --- in nonlinear
QM averages of projectors, $\frac{\langle\psi| r\rangle\langle
r|\psi\rangle}{{\langle\psi|\psi\rangle}}$, may not have an
interpretation in terms of probability (see the section on
eigenvalues).)

It is interesting that the ``collapse"-like property of the
average can be seen also in (\ref{collapselike}) where both
values of ``spin" are averaged separately. Finally, let us
notice that the last observable (\ref{singularlike}) is
unbounded, so that the values of single measurement cannot be
identified with eigenvalues of $\sigma_3$.$\Box$

\subsection{Dynamics}
\label{Dynamics}

Both Kibble and Weinberg assumed that the dynamics of pure
states is given by Hamilton equations with Hamiltonian functions
belonging to the algebra of generalized observables discussed in
Subsec. \ref{Observables}. More generally, all one parameter
flows of canonical transformations are assumed to be integral
curves of the Hamilton equations of motion
$i_X\omega=dH$.

Since the evolution is, in general, nonlinear no dynamical
separation of states and observables is possible (the ordinary
Heisenberg and Schr\"odinger pictures can be formulated only if
the dynamics is linear). Instead, the evolution of observables
is governed by the Poisson bracket resulting from the Hamilton
equations.

The Hamilton equations  in Kibble-Weinberg
theory can be written in a Schr\"odinger-like form
\begin{equation}
i \frac{d}{dt}|\psi\rangle=\hat H(\psi,\bar
\psi)|\psi\rangle\label{nls}
\end{equation}
where $\hat H(\psi,\bar \psi)$ is Hermitian.
 An interesting class of nonlinear Schr\"odinger equations
which are not equivalent to Weinberg's equations,
even though the Hamiltonians are (0,0)-homogeneous, is given by
equations of the form
\begin{equation}
i \frac{d}{dt}|\psi\rangle=\Bigl(\hat H_0+\frac{\langle\psi|\hat
A|\psi\rangle}{{\langle\psi|\psi\rangle}}\hat B-
\frac{\langle\psi|\hat
B|\psi\rangle}{{\langle\psi|\psi\rangle}}\hat
A\Bigr)|\psi\rangle:=
\hat H|\psi\rangle.\label{AB-BA}
\end{equation}
In such a case all the nonlinear terms involving $\hat A$ and
$\hat B$ cancel out in $\langle\psi|\hat
H|\psi\rangle=\langle\psi|\hat H_0|\psi\rangle$, which, for this
reason, cannot be regarded as a Hamiltonian function for
(\ref{AB-BA}).

\begin{ex}\em  Consider a two-level atom irradiated by an
external classical light and assume that the Hamiltonian
function of the atom takes the Weinberg form
\begin{equation}
\langle\psi|\hat H_L+\hat H_{NL}|\psi\rangle
\end{equation}
where the nonlinear $2\times 2$ matrix is
\begin{equation}
\hat H_{NLmn}=\frac{\partial^2 H_{NL}}{\partial\bar
\psi_m\partial\psi_n}= (h_0+\vec h_{NL}\vec \sigma)_{mn}.
\end{equation}
Let $\vec s(\psi,\bar \psi)=\langle\psi|\vec
\sigma|\psi\rangle$. We find
\begin{eqnarray}
\dot {\vec s}&=&\{\vec s,H\}=-i \langle\psi| [\vec \sigma,
\hat H_L+\hat H_{NL}]|\psi\rangle\nonumber\\&=&2(\vec h_L+\vec
h_{NL})\times\vec s= (\vec \omega_L+\vec \omega_{NL})\times\vec
s.
\end{eqnarray}
Defining
\[\left(\begin{array}{c}
\tilde{\omega}_1\\
\tilde{\omega}_2\\
\tilde{\omega}_3
\end{array}
\right)=
\left(\begin{array}{ccc}
\cos\omega t & \sin\omega t & 0 \\
-\sin\omega t & \cos\omega t & 0\\ 0 & 0 & 1
\end{array}\right)
\left(\begin{array}{c}
{\omega}_{NL1}\\ {\omega}_{NL2}\\ {\omega}_{NL3}
\end{array}\right),
\]
with $\omega$ being the frequency of light, we obtain after RWA
the following nonlinear Bloch equations
\begin{eqnarray}
\dot u &=&-\Delta v -\tilde{\omega}_3 v +\tilde{\omega}_2
w\nonumber\\
\dot v &=& \Delta u +\Omega w +\tilde{\omega}_3 u
-\tilde{\omega}_1 w\\
\dot w &=& -\Omega v -\tilde{\omega}_2 u +\tilde{\omega}_1
v\nonumber.
\end{eqnarray}
Choosing $(\tilde{\omega}_1,\tilde{\omega}_2,\tilde{\omega}_3)=
(-(A/2)v,(A/2)u,2\epsilon w)$ we get
\begin{eqnarray}
\dot u &=&-\Delta v -2\epsilon w v +\frac{A}{2}u w\nonumber\\
\dot v &=& \Delta u +\Omega w +2\epsilon w u
-\frac{A}{2}v w\label{Jaynes}\\
\dot w &=& -\Omega v -\frac{A}{2}u^2 -\frac{A}{2}v^2.\nonumber
\end{eqnarray}
Eqs. (\ref{Jaynes}) are identical to Jaynes' neoclassical Bloch
equations \cite{WScully} where $\epsilon$ is the neoclassical
Lamb shift and $A$ is equal to Einstein's coefficient of
spontaneous emission.  So let us check what kind of the
KW-Schr\"odinger equation can lead to the neoclassical
description.

Returning to the non-rotated reference frame we obtain
\begin{equation}
(\omega_{NL1},\omega_{NL2},\omega_{NL3})= (-(A/2)\bar
s_2,(A/2)\bar s_1,2\epsilon \bar s_3)
\end{equation}
and the nonlinear Schr\"odinger equation we are looking for is
\begin{eqnarray}
i \frac{d}{dt}|\psi\rangle&=&\Bigl(\tilde{H}(\psi,\bar \psi){\bf
1}-
\frac{A}{4}\frac{\langle\psi|\sigma_2|\psi\rangle}{{\langle\psi|
\psi\rangle}}\sigma_1+
\frac{A}{4}\frac{\langle\psi|\sigma_1|\psi\rangle}{{\langle\psi|
\psi\rangle}}\sigma_2+
\epsilon\frac{\langle\psi|\sigma_3|\psi\rangle}{{\langle\psi|
\psi\rangle}}\sigma_3\Bigr)|\psi\rangle\nonumber\\
&=&\hat H(\psi,\bar \psi)|\psi\rangle.\label{neo}
\end{eqnarray}
If $\tilde{H}$ is some (0,0)-homogeneous function then the
Hamiltonian is also (0,0)-homogeneous. However
\begin{equation}
\langle\psi|\hat H(\psi,\bar
\psi)|\psi\rangle=\langle\psi|\Bigl(\tilde{H}(\psi,\bar
\psi){ \bf 1}+
\epsilon\frac{\langle\psi|\sigma_3|\psi\rangle}
{{\langle\psi|\psi\rangle}}\sigma_3\Bigr)|\psi\rangle
\end{equation}
cannot play the role of the Hamiltonian function for (\ref{neo})
with $A\neq 0$. For $A=0$ the relevant KW Hamiltonian function
is
\begin{equation}
E{\langle\psi|\psi\rangle} +{1\over
2}\epsilon\frac{\langle\psi|\sigma_3|\psi\rangle^2}
{{\langle\psi|\psi\rangle}}
\end{equation}
and
\begin{equation}
\tilde{H}(\psi,\bar \psi)=E-{1\over 2}\epsilon
\frac{\langle\psi|\sigma_3|\psi\rangle^2}
{{\langle\psi|\psi\rangle}^2}.
\end{equation}$\Box$
\end{ex}

\subsection{Probabilities and Results of Single Measurements}
\label{eigen}

Let $\cal H$ be a finite dimensional Hilbert space and $\hat H$
a Hermitian operator acting in $\cal H$. If $H=\langle\psi|\hat
H|\psi\rangle$ is
the associated observable, the values of single measurements of
$H$ can be defined in at least three equivalent ways. Since in
nonlinear QM the three options will lead to different results we
will use here different names for each of them.

\noindent
a) {\it Eigenvalue\/} $E$ of $H=\langle\psi|\hat H|\psi\rangle$
is the number satisfying for some {\it eigenstate \/} the
equation
\begin{equation}
\hat H|\psi\rangle=\lambda|\psi\rangle
\end{equation}
or
\begin{equation}
\frac{\partial H}{\partial\bar \psi_m}=\lambda\psi_m.\label{eigenvalue}
\end{equation}
b) {\it Diagonal values\/} are solutions of
\begin{equation}
\det(\hat H-\lambda{\bf 1})=0
\end{equation}
or
\begin{equation}
\det\Bigl(\frac{\partial^2 H}{\partial\bar \psi_m\partial
\psi_n}- \lambda\delta_{mn}\Bigr)=0.\label{diagonal value}
\end{equation}
c) The third possibility comes from the fact that, since
eigenstates form a complete orthogonal set of vectors
in $\cal H$, any solution of the Schr\"odinger equation $i
\frac{d}{dt}|\psi\rangle=\hat H|\psi\rangle$
can be expressed as
\begin{equation}
\left(\begin{array}{c}
\psi_1(t)\\
\vdots\\
\psi_N(t)
\end{array}\right)=
\left(\begin{array}{c}
\psi_1(0)e^{-i \omega_1t}\\
\vdots\\
\psi_N(0)e^{-i \omega_Nt}
\end{array}\right)
\end{equation}
and the frequencies $\omega_n$ can be termed the {\it
eigenfrequencies\/}. The three possibilities can be used for
definitions of the results of single measurements also in
nonlinear QM.

Definitions (\ref{eigenvalue}) and (\ref{diagonal value}) do not
explicitly use the bilinearity of $H$, so can be naturally
adapted also in the nonlinear framework. It is interesting that
then they are not equivalent.

\begin{ex}\em  The so-called ``simplest nonlinearity"
considered in experiments designed as tests of Weinberg's
nonlinear QM corresponds to the following Hamiltonian function
\begin{equation}
\langle\psi| \hat
H_0|\psi\rangle+\epsilon\frac{\langle\psi|\sigma_3|
\psi\rangle^2}{{\langle\psi|\psi\rangle}}
\end{equation}
where $\hat H_0={E_1,0 \choose 0, E_2}$.  Denote $\psi_1=\sqrt
{A_1}e^{i \alpha_1},\, \psi_2=\sqrt {A_2}e^{i \alpha_2}$. The
eigenvalue condition reads
\begin{eqnarray}
\Bigl\{ E_1 +\epsilon\Bigl[2{A_1 - A_2\over A_1+A_2}-\Bigl(
{A_1 - A_2\over A_1+A_2}\Bigr)^2\Bigr]\Bigr\} \sqrt{A_1}
&=&\lambda \sqrt{A_1}\\
\Bigl\{ E_2 +\epsilon\Bigl[-2{A_1 - A_2\over A_1+A_2}-\Bigl(
{A_1 - A_2\over A_1+A_2}\Bigr)^2\Bigr]\Bigr\} \sqrt{A_2}
&=&\lambda \sqrt{A_2}.
\label{Ex2.2}
\end{eqnarray}
For $A_2=0$ we find $\lambda_+=E_1+\epsilon$, for $A_1=0$
$\lambda_-= E_2+\epsilon$. If $A_1\ne 0$ and $A_2\ne 0$ we get,
for $\epsilon\ne 0$,
\[
{A_1 - A_2\over A_1+A_2}={E_2-E_1\over 4\epsilon}
\]
which implies ($A_1$ and $A_2$ are positive)
\begin{equation}
|E_2-E_1|<4|\epsilon|\label{..<4eps}
\end{equation}
and
\[
\lambda_0 ={1\over 2}\Bigl( E_1+E_2 -{1\over
8\epsilon}(E_1-E_2)^2\Bigr)
\]
with the eigenvector
\begin{equation}
|\psi\rangle =
\left(
\begin{array}{c}
\bigl(1+{E_1-E_2\over 4\epsilon}\bigr)^{-1/2}e^{i \alpha_1}\\
\bigl(1-{E_1-E_2\over 4\epsilon}\bigr)^{-1/2}e^{i
\alpha_2}\label{Ex2.2eigen}
\end{array}\right)
\end{equation}
We find three eigenvalues although the dimension of the Hilbert
space is two. The third eigenstate exists only for $\epsilon$
satisfying the condition (\ref{..<4eps}).

Of course, all the eigenstates are stationary. The eigenstate
corresponding to $\lambda_0$ is not orthogonal to the remaining
two eigenstates, which excludes the ordinary probability
interpretation in terms of projectors.

Let us now calculate the diagonal values of this Hamiltonian
function. The nonlinear Hamiltonian is
\begin{equation}
\hat H=
\left(\begin{array}{cc}
E_1+\epsilon(8p^3-20p^2+16p-3) &
8\epsilon|\psi_1|^2|\psi_2|^2\psi_1\bar
\psi_2\\
8\epsilon|\psi_1|^2|\psi_2|^2\bar \psi_1\psi_2 &
E_2+\epsilon(-p^3 +4p^2+1)
\end{array}\right)
\end{equation}
where the state is assumed normalized and $p=|\psi_1|^2$. The
matrix is Hermitian and its eigenvalues (=diagonal values of
$H$) are
\begin{eqnarray}
E_\pm &=& {1\over
2}\Bigl(E_1+E_2-2\epsilon(8p^2-8p+1)\nonumber\\
&\,&\pm\Bigl[(E_1-E_2)^2-8\epsilon\bigl\{(E_1-E_2)
(4p^3-6p^2+4p-1)\nonumber\\
&\,\,&+2\epsilon(20p^4-40p^3+28p^2-8p+1)\bigr\}\Bigr]^{1/2}
\Bigr).
\end{eqnarray}
The diagonal values, in opposition to eigenvalues,
 are in the nonlinear case {\it functions\/}
  and the number of
them, again as opposed to eigenvalues, is always equal to the
dimension of the suitable Hilbert
space.

The solution of the respective nonlinear Schr\"odinger equation
is
\begin{equation}
\left(\begin{array}{c}
\psi_1(t)\\
\psi_2(t)
\end{array}\right)=
\left(\begin{array}{c}
\psi_1(0)e^{-i
(E_1+2\epsilon\langle\sigma_3\rangle-
\epsilon\langle\sigma_3\rangle^2)t}\\
\psi_2(0)e^{-i (E_2-2\epsilon\langle
\sigma_3\rangle-\epsilon\langle\sigma_3\rangle^2)t}
\end{array}\right)
\end{equation}
where the averages in the exponents
 are integrals of motion. It follows that the
eigenfrequencies are also state dependent functions but differ
from the diagonal values.
\end{ex}

In linear QM mechanics observables correspond to averages hence,
at experimental level, to random variables. With any random
variable one can associate its higher moments. The higher
moments, on the other hand, can be used to deduce a probability
interpretation of the theory. Therefore, one of the essential
points of any generalized algebra of observables is the question
of the representability of higher moments corresponding to some
given observable. In linear QM the problem is solved by the
spectral theorem.

The first interesting result concerning the nonlinear {\it
eigenvalues\/}, noticed by Weinberg in \cite{W2}, is the
following

\begin{lem} Let $F$ and $G$ be two (1,1)-homogeneous
observables possessing a common eigenstate $|\psi\rangle$, i.e.
$
\frac{\partial F}{\partial\bar \psi_{k}}=
f\psi_k
$
and c.c.,
$
\frac{\partial G}{\partial\bar \psi_{k}}=
g\psi_k
$
and c.c.,
and let
$
F*G=\frac{\partial F}{\partial\psi_{k}}
\frac{\partial G}{\partial\bar \psi_{k}}.
$
Then
$
\frac{\partial (F*G)}{\partial\bar \psi_{k}}=
fg\psi_k
$ and c.c.
\end{lem}

The homogeneity implies that an average of $F$ in an eigenstate
is equal to the respective eigenvalue. For finite dimensional
Hilbert spaces we know also that a number of eigenvalues is not
smaller than the dimension of the space (a result from the Morse
theory) and  the eigenvalues are critical points of
averages. For averages defined on the whole Hilbert/projective
space their maxima and minima must be found at critical points
(averages defined on a finite dimensional projective space are
smooth functions defined on a compact set).  These facts suggest
that the eigenvalues are correct candidates for the results of
single measurements of the generalized observables. However, the
following examples show that the problem is not that simple.

\begin{ex}\em  The observable
\begin{equation}
A=\frac{{\langle\psi|\psi\rangle}^2}
{\langle\psi|\sigma_3|\psi\rangle}
\end{equation}
satisfies the Weinberg's homogeneity  requirements. Its eigenvalues are $\pm 1$
hence
do not bound the averages. It follows that such singular
observables must be excluded if we want to have the probability
interpretation in terms of the eigenvalues. The algebra of
observables would have to be restricted to (1,1)-homogeneous
smooth functions defined of the {\it whole\/} Hilbert/projective
space.  This requirement is not very restrictive. Notice that no
difficulties will appear if we will apply to $A$ the
interpretation in terms of diagonal values or
eigenfrequencies.$\Box$\end{ex}

\begin{ex}\em  In QM we often encounter observables whose
eigenstates are degenerate. Consider now
\begin{equation}
H=E{\langle\psi|\psi\rangle}+\epsilon\frac{\langle\psi|
\sigma_3|\psi\rangle^3}{{\langle\psi|\psi\rangle}^2}.
\end{equation}
Its eigenvalues are $E\pm\epsilon$ and $E$. The number of them
is hence greater than the dimension of the Hilbert space no
matter how small $\epsilon$ is. Again, no problems appear if we
apply the diagonal values or eigenfrequencies interpretation.
Had we substituted the first term with some ${E_1,0 \choose 0,
E_2}$, where $E_1\neq E_2$, we would have obtained the third
eigenvalue provided $|E_1-E_2|\leq 6|\epsilon|$ so that the
eigenvalue will not appear for sufficiently small $\epsilon$.
This phenomenon seems to be related to the
Kolmogorov-Arnold-Moser theorem where a dimension of an
invariant torus may not change with nonlinear perturbation if a
nonperturbed Hamiltonian system is nondegenerate and the
perturbation is not too large \cite{Arnold}. $\Box$\end{ex}

\begin{ex}\em  Let
\begin{equation}
H=\langle\psi|\hat H_0|\psi\rangle
+\epsilon\frac{\langle\psi|\sigma_3|\psi\rangle^{2N}}
{{\langle\psi|\psi\rangle}^{2N-1}}
\end{equation}
where $\hat H_0={E_1,0 \choose 0, E_2}$. Two eigenvalues
corresponding to the eigenstates $| \pm\rangle$ of $\sigma_3$ are
$E_+=E_1+\epsilon$ and $E_-=E_2+\epsilon$ so are shifted by the
same amount.  The third eigenstate occurs if
\begin{equation}
|\epsilon|>\frac{|E_1-E_2|}{4N}.
\end{equation}
The third eigenvalue
\begin{equation}
E_0={1\over
2}(E_1+E_2)+\epsilon(1-2N)\Bigl(\frac{E_2-E_1}{4N\epsilon}\Bigr)^
{\frac{2N}{2N-1}}
\end{equation}
tends to $E_1$ with $N\to \infty$.
This example is important. It shows that the ordinary
projection postulate cannot be consistently applied, for
calculation of probabilities coresponding to nonlinear {\em
eigenstates\/}, in experiments whose goal is
elimination of {\em all\/} nonlinear corrections to QM.
  $\Box$\end{ex}

\begin{ex}\em  Consider
\begin{equation}
H=E{\langle\psi|\psi\rangle}+\epsilon\frac{\langle\psi|\sigma_3|
\psi\rangle^2}{{\langle\psi|\psi\rangle}}.
\end{equation}
The eigenvalues are $E$ and $E+\epsilon$. Probabilities
calculated by means of
\begin{equation}
\langle H\rangle=(E+\epsilon) p_{E+\epsilon}+Ep_E
\end{equation}
and the normalization of probability are
\begin{equation}
p_{E+\epsilon}=\langle \sigma_3\rangle^2\qquad p_E=1-\langle
\sigma_3\rangle^2 .
\end{equation}
We know that $H*H$ has eigenvalues $(E+\epsilon)^2$ and $E^2$,
so we can calculate the probabilities by means of $\langle
H*H\rangle=(E+\epsilon)^2 p_{E+\epsilon}+E^2 p_E$ and the
normalization condition. Since
\begin{equation}
\langle H*H\rangle=E^2 +(4\epsilon^2 +2E\epsilon)\langle
\sigma_3\rangle^2-
3\epsilon^2 \langle \sigma_3\rangle^4
\end{equation}
we obtain
\begin{equation}
p_{E+\epsilon}=\frac{(4\epsilon^2 +2E\epsilon)\langle
\sigma_3\rangle^2-
3\epsilon^2 \langle \sigma_3\rangle^4}{2E\epsilon+\epsilon^2}
\end{equation}
which is, of course, inconsistent with the previous result.  The
troubles arise because of the nonassociativity of $*$. An
analogous result was derived by Jordan \cite{Jor}, who showed
that propositions cannot be defined consistently within the
$*$-algebraic approach to Weinberg's observables. We conclude
that $*$ cannot be applied for a definition of higher moments
and the above lemma is useless from the viewpoint of the
probability interpretation. $\Box$ \end{ex}

Next candidate for a single result of a measurement of some $H$
is the diagonal value. We have already seen that the diagonal
values are state dependent functions. There is no general
guarantee that a diagonal value of an integral of motion is
itself an integral of motion. In fact, as the Weinberg equations are
not necessarily integrable for ${\rm dim}{\cal H}>2$, whereas
the number of the diagonal values is equal to the dimension of
$\cal H$, the diagonal values, for $N>2$, can be time
independent only for integrable systems.  In addition, as
diagonal values are roots of algebraic equations, they even do
not have to be differentiable functions of states. Therefore, we
cannot find for them any general equation of motion.

In spite of these disadvantages the diagonal values are in some
respects superior to eigenvalues.  If for two observables $A$
and $B$ the commutator $\hat A\hat B-\hat B\hat A=0$, then $\hat
A$ and $\hat B$ can be diagonalized simultaneously and the
product $\hat A\hat B$ has eigenvalues  being products of
the eigenvalues of $\hat A$ and $\hat B$. This leads to the
following unique probability interpretation.  Let $U(\psi,\bar
\psi)$ be the unitary transformation
diagonalizing $\hat H$. Then
\begin{equation}
H=\sum_nE_n(\psi,\bar \psi)|(U\psi)_n|^2.
\end{equation}
The functions $p_n=|(U\psi)_n|^2/{\langle\psi|\psi\rangle}$ are
probabilities resulting from the higher moments procedure.

Let us recall that for two observables $F*G=F\bar * G$. The
nonassociativity of $*$ is reflected in the $\bar *$-algebra of
observables in the following non-uniqueness of $\bar *$.
Although always
\begin{equation}
\langle\psi|\hat{(H\bar *\dots\bar *
H)}|\psi\rangle=\langle\psi|\hat H\dots\hat H|\psi\rangle
\end{equation}
but for non-bilinear observables
\begin{equation}
\hat{(H\bar *\dots\bar * H)}\neq\hat H\cdot\dots\cdot\hat H.
\end{equation}
The eigenfrequencies approach can be applied only to systems
that are close to integrable \cite{Arnold}. For finite
dimensional Hamiltonian systems some general KAM theorems are
known. It is known that components of state vectors exhibit in
such a case, unless some resonant initial conditions occur, a
quasi periodicity described by \cite{W2}
\begin{equation}
\psi_k(t)=\sum_{n_1\dots n_N}c_k(n_1\dots n_N)
e^{-i \sum_\nu n_\nu\omega_\nu t}\label{int}
\end{equation}
where both the amplitudes and the frequencies are dependent on
initial conditions, as shown in our first example in this
subsection, and the sums run over all positive and negative
integers.  The homogeneity condition implies that
\begin{equation}
{\langle\psi|\psi\rangle}=\sum_k\sum_{n_1\dots}|c_k(n_1\dots)|^2
\end{equation}
and
\begin{equation}
H=\sum_k\sum_{n_1\dots}|c_k(n_1\dots)|^2\sum_\nu
n_\nu\omega_\nu.
\end{equation}
These formulas lead to the natural interpretation of the
eigenfrequencies $\sum_\nu n_\nu\omega_\nu$ as the results of
single measurements and the respective normalized coefficients
as probabilities. The interpretation can be naturally extended
to other observables by assuming that all observables generate
one parameter groups of canonical transformations {\it via\/}
the Hamilton equations of motion. The same result would be also
obtained by means of the ``standard" theory of measurement where
we assume that the so-called pre-measurement is described by an
interaction Hamiltonian function which is proportional to the
observable measured, and that all the measuring procedures are
based on measurements of observables that are bilinear like in
linear QM. To be precise, we must remark that this procedure
(chosen finally by Weinberg) is not unique either because there
is no unique description of the composite ``object+observer"
system even if the observer is assumed to be linear (see remarks
in the section devoted to observables).  Another difficulty
 is that the assumption about measurements based
only on bilinear observables cannot be formulated in a
relativistic way. In standard measurements one measures
positions or momenta. A theory where momenta are always linear
observables
whereas the Hamiltonian function is a nonlinear observable
cannot be relativistically covariant. Perhaps it was this point
 that made Weinberg admit that he ``could not find
any way to extend the nonlinear version of quantum mechanics to
theories based on Einstein's special theory of
relativity"~\cite{Dreams}.
My own proposal presented in part II avoids this problem,
because I introduce the nonlinearity through {\em entropy\/},
which is not an observable, hence without any modification of
the algebra of observables.
To better understand mutual relations between eigenvalues and
eigenfrequencies it is interesting to compare the two notions in
situations where a number of eigenvalues is different from this
of eigenfrequencies.

\begin{ex}\em  We know that
\begin{equation}
H=E{\langle\psi|\psi\rangle}+\epsilon\frac{\langle\psi|
\sigma_3|\psi\rangle^3}{{\langle\psi|\psi\rangle}^2}.
\end{equation}
has three eigenvalues $E\pm\epsilon$ and $E$. The
eigenfrequencies are
\begin{eqnarray}
E_1(\psi,\bar \psi)&=&E+\epsilon(3\langle\sigma_3\rangle^2-
2\langle\sigma_3\rangle^3)\\ E_2(\psi,\bar
\psi)&=&E+\epsilon(-3\langle\sigma_3\rangle^2-
2\langle\sigma_3\rangle^3)
\end{eqnarray}
The three eigenvalues correspond to eigenvectors satisfying
\[
\begin{array}{c}
\\
\langle\sigma_3\rangle\\
\,
\end{array}=
\left\{\begin{array}{rl}
0 & {\rm for}\quad \psi^{0}\\ 1 & {\rm for}\quad \psi^{1}\\ -1 &
{\rm for}\quad \psi^{-1}
\end{array}\right.
\]
In this notation
\[
\begin{array}{rll}
E_1(\psi^0,\bar \psi^0)=& E &\mbox{ with probability
$p_1(\psi^0,\bar \psi^0)=1/2$}\\ E_2(\psi^0,\bar \psi^0)=& E
&\mbox{ with probability $ p_2(\psi^0,\bar
\psi^0)=1/2$}\\
E_1(\psi^1,\bar \psi^1)=& E+\epsilon &\mbox{ with probability
$p_1(\psi^1,\bar \psi^1) =1$}\\ E_2(\psi^1,\bar \psi^1)=&
E-5\epsilon &\mbox{ with probability $p_2(\psi^1,\bar \psi^1)
=0$}\\ E_{1}(\psi^{-1},\bar \psi^{-1})=& E+5\epsilon &\mbox{
with probability $p_{1}(\psi^{-1},\bar \psi^{-1}) =0$}\\
E_2(\psi^{-1},\bar \psi^{-1})=& E-\epsilon &\mbox{ with
probability $p_2(\psi^{-1},\bar \psi^{-1}) =1$}.
\end{array}\]
As expected even in an eigenstate there are two
eigenfrequencies.  The ones which are not equal to the
eigenvalues occur with probabilities 0.$\Box$\end{ex}

A reader of the main Weinberg's paper may be a little bit
confused with what is finally understood as a result of a single
measurement. A half of the paper suggests that this role will be
played by eigenvalues, then probabilities are defined in terms
of eigenfrequencies and, finally, in the analysis of a two-level
atom a difference of eigenvalues is treated as the energy
difference while the difference of the eigenfrequencies is
treated as the frequency of the emitted photon. It seems that
from the viewpoint of the analysis above the eigenfrequency
difference should be treated as the energy difference of atomic
levels. This is one of the subtleties that have been omitted in
the analysis of experiments.

I think that each of the three possibilities lacks elegance and
generality, and this is one of the reasons one should look for a
possibility of introducing nonlinearities without altering
observables.

\section{Composite Systems and Nonlocality}
\label{COMPOSITE}

One of the most often quoted ``impossibility theorems" about
nonlinear QM is that {\it any\/} such theory must imply faster
than light communications. This statement is evidently too
strong. In fact, we will
show in part II that there exists a rich class of theories where the
mentioned phenomenon does not occur. As we shall see the
theories must satisfy two conditions.

\begin{enumerate}
\item  Their interpretation must not be based on the ``collapse of a
state vector" postulate. It means that any reasoning based on
this postulate has to be regarded as unphysical. (In linear QM
there exist such limitations. For example, all ``counterfactual"
problems like EPR paradox or Bell theorem can be eliminated
trivially by rejecting reasonings involving alternative
measurements.) Interpretations of QM that do not introduce the
collapse (projection) postulate exist, to mention the many
worlds one
\cite{MWI}.
%,\cite{decoherent}

\item Observables corresponding to subsystems must be functionals
depending on {\it density matrices of those subsystems\/} (the
Polchinski postulate \cite{Polchinski}).
\end{enumerate}

In the second part of the paper I shall prove some general,
quite strong
theorems related to the latter condition. Here, we shall limit
ourselves to several simple examples showing different ways of
making quantum mechanical nonlocality ``malignant", to use the
marvelous phrase of Bogdan Mielnik.

It seems it was Nicolas Gisin who was the first to observe that
a nonlinear evolution can lead to a faster-than-light
communication between two separated systems. His argument was
the following
\cite{Gisin1}. Let $\cal H$ be a finite dimensional Hilbert
space, $| {\psi_i}\rangle,\,| {\phi_j}\rangle\in \cal H$, $i=1,\dots,n$,
$j=1,\dots,m$,
$\langle{\psi_i}|{\psi_j}\rangle=\langle{\phi_i}|{\phi_j}
\rangle=\delta_{ij}$.
Then the following lemma holds.

\begin{lem} If for some nonvanishing probabilities
$x_i,\,y_j$
\begin{equation}
\sum_i x_i| {\psi_i}\rangle\langle {\psi_i}|=
\sum_j y_j| {\phi_j}\rangle\langle {\phi_j}|
\end{equation}
then there exist two orthonormal bases $\{| {\alpha_i}\rangle\}$,
$\{| {\beta_j}\rangle\}$ in some Hilbert space $\cal H'$ and the
state
\begin{equation}
| \chi\rangle =\sum_i\sqrt{x_i}| {\psi_i}\rangle\otimes|
{\alpha_i}\rangle=
\sum_j\sqrt{y_j}| {\phi_j}\rangle\otimes|
{\beta_j}\rangle.
\end{equation}
$\Box$\end{lem} The proof can be found in \cite{Gisin1}. The
meaning of the lemma is that the two decompositions of the
density matrix can be obtained by means of the EPR correlations.
Indeed, we can take the density matrix $| \chi\rangle\langle \chi|$ and
trace out $\cal H'$. The resulting density matrices are these
appearing in the lemma. Assume now that we have a nonlinear
evolution of pure states
\begin{equation}
| {\psi_i}\rangle\langle {\psi_i}|\mapsto g_t(| {\psi_i}\rangle\langle
{\psi_i}|). \label{Gis}
\end{equation}
Then, in general,
\begin{equation}
\sum_i x_i g_t(| {\psi_i}\rangle\langle {\psi_i}|)\neq
\sum_j y_j g_t(| {\phi_j}\rangle\langle {\phi_j}|)\label{Pos}
\end{equation}
even if initially the two decompositions were equal. The
important assumption leading to the faster-than-light
communication is that each of the pure state sub-ensembles
evolves according to (\ref{Gis}), even if the whole state is
given by their convex combination. This seems reasonable if we
assume that the ensemble consists of the collapsed
sub-ensembles.  If we apply some ``no-collapse" interpretation
then the argument cannot be consistently formulated. However, it
has to be stressed that the form of the evolution appearing in
(\ref{Pos}) can be derived from quite general assumptions. It
was shown in a rigorous way in the language of theory of cathegories
 by Posiewnik \cite{Pos} that this form is implied by
Mielnik's definition of mixed states as probability measures on
the set of pure states. This remark has quite nontrivial
consequences: The description of mixed states in terms of
probability measures (Mielnik's ``convex approach") combined
with putative nonlinearity of the Schr\"odinger equation leads
to faster that light communication.  If one wants to get rid of
such difficulties one cannot keep the figure of states convex.

Gisin, in his second paper \cite{Gisin2}
considered an example of a nonlinear evolution taken from the
Weinberg paper \cite{W2}. Consider an ensemble of pairs of
spin-1/2 particles in the singlet state and assume that in one
arm of the experiment the evolution is given by the Hamiltonian
function
$
{\langle\psi|\sigma_3|\psi\rangle^2}/{{\langle\psi|
\psi\rangle}}.
$
An experimenter in the other arm chooses between two settings of
his Stern-Gerlach device and decomposes, by means of the EPR
correlations, the ensemble in the other arm into subensembles
corresponding to, say, spins up or down in the $z$ direction, or
spins up or down in some $u$ direction tilted at $45^\circ$ with
respect to the $z$-axis. The evolution equation, up to an
overall phase, is given by
$
i
\frac{d}{dt}|\psi\rangle=2\langle\sigma_3\rangle\sigma_3|
\psi\rangle.
$
If the initial state is either up or down in the $z$ direction
the evolution is stationary and the average of $\sigma_2$ is
always zero. If the state is either up or down in the $u$
direction the spin is precessing around the $z$ axis but the
sense of the rotation is opposite for the up and down states.
After a time of a quarter of the ``Larmor period" the spin will
have the same positive value of $\langle\sigma_2\rangle$. And
this result can be detected by an observer in the arm with the
nonlinearity.

As we can see Gisin in both of his proofs ignored
details of the description and the evolution of the ``large"
system. To be precise he should rather define a Hamiltonian
function of the two arms, then solve the nonlinear Schr\"odinger
equation for the whole system or calculate the evolution of the
average of $\sigma_2$ in one arm. This is the crucial point.  We
know already that the description of composite systems is not
unique as long as we know only the evolution of pure states of
subsystems. It will be shown below that the Gisin's telegraph
will never work if we describe the whole system in  the
way proposed by Pochinski. It will be shown also that the proof
{\it is\/} valid for
the specific choice of the description proposed by Weinberg.

But before we shall pass on to the details of the calculations
let us consider another ``general proof" which was formulated by
myself independently and simultaneously with this of Gisin
\cite{mobility}.
Consider a polarizing Mach-Zehnder whose first beam splitter is
an analyzer of circular polarizations (for a detailed
description of the polarizing Mach-Zehnder interferometer see
\cite{bell}).
A reflected beam is righthanded and the transmitted one ---
lefthanded. The lefthanded beam passes through a half-wave plate
and then the two beams are phase shifted by $\alpha$ with
respect to each other and recombined.

If a source produces a linear polarization state, say
\begin{equation}
|\psi\rangle ={1\over\sqrt2}(| +\rangle +|-\rangle
).\label{linear}
\end{equation}
the interferometer transforms the whole state into

\begin{equation}
|\psi'\rangle ={1\over2}\bigl(i (e^{i \alpha}+1)| +\rangle
+(1-e^{i \alpha})| +\rangle \bigr).\label{linear'}
\end{equation}
and we observe the interference between the two outgoing
channels.

If, however, the source produces a singlet state \cite{PZuk}
\begin{equation}
|\phi\rangle ={1\over\sqrt2}(| +\rangle | +\rangle +| -\rangle
|-\rangle ),\label{singlet}
\end{equation}
 the whole state transforms as follows
\begin{equation}
\begin{array}{rlll}
|\phi'\rangle = {1 \over 2}( ie^{i\alpha}&| +\rangle |+\rangle &
- e^{i\alpha}&|+\rangle |+\rangle \\ +i &|-\rangle |+\rangle &+
& | -\rangle |+\rangle ).\label{singlet'} \end{array}
\end{equation}
and there cannot appear the interference in the interferometer
(subsystem {\bf I}):{ \it The linear dependence of the states
which interfered in the linear polarization case is destroyed by
orthogonality of their singlet state partners\/}. The lack of
the interference in {\bf I} is therefore a {\it nonlocal\/}
phenomenon.

We know that nonlinear evolutions in a Hilbert space do not
conserve scalar products (the ``mobility phenomenon"
\cite{MielnikMob}) and states that are initially orthogonal may
loose their orthogonality during the course of the evolution. It
is clear that if we will violate in some system {\bf II}
separated from {\bf I} the orthogonality
of $| {+}\rangle$ and $| {-}\rangle$ because of some nonlinearity then
the photons in {\bf I} will start to interfere and the
interference will be the stronger the more violation of the
orthogonality has been obtained. Putting it differently, tracing
over a space where the evolution does not conserve scalar
products may result in some ``remains" of the traced out system
in the reduced density matrix describing the remote, separated
system.

This argument looks general but a careful reader has probably
noticed that one additional assumption has been smuggled here:
It is implicitly assumed that a nonlinear evolution in $\cal H$
can be extended to $\cal H\otimes \cal H'$ in such a way that a
solution describing the composite system is of the form
\begin{equation}
| \phi\rangle\otimes| +\rangle +| {\phi'}\rangle\otimes| -\rangle
\end{equation}
where $| \phi\rangle$ and $| {\phi'}\rangle$ are solutions of the
subsystem's nonlinear evolution equation that are ``in
mobility", that is whose scalar product is not conserved. This
assumption is a strong limitation. In fact, it never occurs if
separated systems are described in a Polchinski's way. However, the
proposal of Weinberg {\it does allow\/} for such pathological
solutions.

In what follows we shall formulate the arguments of Gisin and
myself in a precise way within the framework of Weinberg's
approach.

Let us assume that we have two separated systems, {\bf I} and
{\bf II}, described by Hamiltonian functions
\begin{equation}
H_1(\varphi,\bar \varphi)=E_1\langle \varphi|\varphi\rangle ,
\qquad
H_2(\chi,\bar \chi)=E_2\langle \chi|\chi\rangle
+\epsilon{\langle \chi|\sigma_3|\chi\rangle ^2\over\langle
\chi|\chi\rangle }. \label{IandII}
\end{equation}
According to Weinberg, the Hamiltonian function of the whole
{\bf I}+{\bf II} system is given by
\begin{equation}
H(\psi,\bar \psi)=
\sum_lH_1(\varphi_{(l)},\bar \varphi_{(l)})+
\sum_kH_2(\chi_{(k)},\bar \chi_{(k)}),\label{Wcomposite}
\end{equation}
where $\varphi_{k(l)}=\chi_{l(k)}=\psi_{kl}$.

A general solution of the nonlinear Schr\"odinger equation
corresponding to (\ref{Wcomposite}) takes the required form
\begin{equation}
|\psi\rangle=
| {\xi_1}\rangle\otimes| {\phi}\rangle+| {\xi_2}\rangle\otimes|
\psi\rangle\label{gensol}
\end{equation}
where $| {\xi_1}\rangle={\alpha \choose 0}, | {\xi_2}\rangle={0\choose
\alpha },
\alpha =\exp(-i  E_1t)$   and $| \phi\rangle={x\choose y},
|\psi\rangle={x'\choose y'}$ are some solutions of the single
particle nonlinear Schr\"odinger equation.

The Hamiltonian functions
\begin{equation}
H_1(\psi, \psi)=\sum_lH_1(\varphi_{(l)},\bar \varphi_{(l)})
\quad {\rm and}
\quad H_2(\psi,\bar \psi)=\sum_kH_2(\chi_{(k)},\bar \chi_{(k)})
\end{equation}
commute because $H_1(\psi,\bar \psi)=E_1\langle \psi,\psi\rangle
$. Bearing in mind that there is no interaction part in $H$, we
conclude that there is no flow of energy between {\bf I} and
{\bf II}.

Let us now calculate the reduced density matrix of the {\it
linear\/} subsystem {\bf I}. With the notation of
(\ref{gensol}), we find
$
\rho_1={1\over 2}(|\xi_1\rangle \langle \xi_1|+
| \xi_2\rangle \langle \xi_2| +\langle
\psi|\phi\rangle |\xi_1\rangle \langle \xi_2|+
\langle \phi|\psi\rangle |\xi_2\rangle \langle \xi_1|).
$
The off-diagonal elements vanish in the linear theory
$(\epsilon=0)$ if initially the states are orthogonal, and we
get a ``fully mixed" state.  For $\epsilon\neq 0$ these
coherences oscillate with the mobility frequency.  There
therefore exist observables whose average values oscillate in
this way. For example, the components of spin satisfy $\langle
\sigma_1\rangle =\langle
\sigma_3\rangle =0$, but $
\langle \sigma_2\rangle$
is proportional to $\sin\bigl(4\epsilon\langle\sigma_3\rangle
t\bigr)$ hence depends on the parameter characterizing the
nonlinearity of the other, separated and nonintercating system.

Let us complete the analysis of this part with two comments.

\begin{enumerate}
\item  The telegraph is based on the fact that an average of an
observable in the linear system {\bf I} depends on parameters of
the Hamiltonian function of the nonlinear system {\bf II}.
Recalling the form of the Poisson bracket equation for
observables $A_1$ related to {\bf I}
$
\dot{A_1}=\{A_1, H_1+H_2\}
$
we can see that such a dependence is possible if and only if
$
\{A_1,H_2\}\neq 0.
$
It is an easy exercise to check that this condition is met
indeed in the case of $A= \langle\psi|\sigma_2|\psi\rangle$ and
$H_2$ from (\ref{Wcomposite}). It follows that the necessary an
sufficient condition for elimination of telegraphs based on the
mobility phenomenon is that {\it any observables corresponding
to separated systems must be in involution\/} with respect to
the Poisson bracket generating the evolution of observables. The
observation that such observables are not necessarily in
involution in the Weinberg approach is due to J.~Polchinski
\cite{Polchinski}.

\item If one of the systems is linear, then the telegraph based on
the mobility phenomenon can be used for sending information only
from the nonlinear system to the linear one. In Gisin's
telegraph one utilizes the remote preparation of mixtures
entering the nonlinear system. Accordingly, this kind of
telegraph works in the opposite direction hence cannot be
equivalent to this based on the mobility phenomenon.
\end{enumerate}

Let us describe now the Gisin's telegraph more precisely
\cite{Cesena}.
Consider again the same Hamiltonian functions of the subsystems
and the Weinberg's description of the whole one (Eqs.
(\ref{IandII}) and (\ref{Wcomposite})).

The sender is free to choose a basis in his Hilbert space by a
rotation of his Stern-Gerlach device. Let the unitary matrix
with unit determinant
\begin{equation}
\left(\begin{array}{cc}
\alpha &\beta\\
-\bar \beta&\bar \alpha
\end{array}
\right)
\end{equation}
describe the freedom in the choice of bases in the {\it
linear\/} subsystem and the relation between the components of
the whole state in the chosen basis and in the spin up-down one
is given by
\begin{equation}
\left(\begin{array}{cc}
\psi_{11} &\psi_{12}\\
\psi_{21}&\psi_{22}
\end{array}
\right)
=
\left(\begin{array}{cc}
\alpha &\beta\\
-\bar \beta&\bar \alpha
\end{array}
\right)
\left(\begin{array}{cc}
\psi_{++} &\psi_{+-}\\
\psi_{-+}&\psi_{--}
\end{array}\right).
\end{equation}
The state is initially the singlet, which means that
\begin{equation}
\left(\begin{array}{cc}
\psi_{11}(0) &\psi_{12}(0)\\
\psi_{21}(0)&\psi_{22}(0)
\end{array}
\right)
= {1\over \sqrt 2}\left(\begin{array}{cc}
\alpha &\beta\\
-\bar \beta&\bar \alpha
\end{array}
\right)
\left(\begin{array}{cc}
0 &1\\ -1&0
\end{array}\right)
= {1\over \sqrt 2}\left(\begin{array}{cc} -\beta &\alpha\\ -\bar
\alpha&-\bar \beta
\end{array}
\right).
\end{equation}
Finally, the solution for the whole system is
\begin{equation}
|\psi\rangle ={1\over \sqrt 2}
e^{-i (E_1+E_2-\epsilon X^2)t}
\left(\begin{array}{cc}
-\beta e^{-i 2\epsilon Xt} & \alpha e^{i 2\epsilon Xt}\\ -\bar
\alpha e^{i  2\epsilon Xt} & - \bar \beta e^{-i  2\epsilon Xt}
\end{array}\right)
\end{equation}
where $X=|\beta|^2-|\alpha|^2$.  The reduced density matrix of
the {\it nonlinear\/} system {\bf II} reads
\begin{equation}
\rho^{II}={1\over 2} {\bf 1}+{\rm Re} (\bar \alpha\beta)
\sin \bigl(4\epsilon(|\alpha|^2-|\beta|^2)t\bigr)\sigma_2+
{\rm Im} (\bar \alpha\beta)
\sin \bigl(4\epsilon(|\alpha|^2-|\beta|^2)t\bigr)\sigma_1.
\end{equation}
The average of $\sigma_2$ in the nonlinear system is
\begin{equation}
\langle\sigma_2\rangle=2{\rm Re} (\bar \alpha\beta)
\sin \bigl(4\epsilon(|\alpha|^2-|\beta|^2)t\bigr)
\end{equation}
hence depends on the choice of basis made in the linear one.

Notice that since the Hamiltonian function of the linear
subsystem is proportional to ${\langle\psi|\psi\rangle}$ it is
in involution with any observable.  In particular
$
\{\langle\sigma_2\rangle,H_1\}=0
$
which means that the commutability of observables corresponding
to separated systems {\it is not\/} a sufficient condition for
the nonexistence of faster-than-light telegraphs.  The
dependence of $\langle\sigma_2\rangle$ on $\alpha$ and $\beta$
follows from the dependence of $H_2$ on these parameters. We
have remarked already in the section dealing with observables
that the Weinberg's choice is basis dependent: By a change of
basis in {\bf I} we can change a value of energy in {\bf II}.

It becomes clear now under what conditions this kind of
pathology can be eliminated. Consider a density matrix $\rho$
describing a ``large" system. A change of basis in a subsystem
{\bf I} is represented by the unitary transformation
\begin{equation}
\rho\mapsto U_{I}\otimes{\bf 1}_{II}\rho U_{I}^{-1}\otimes{\bf
1}_{II}.
\end{equation}
Any function of $\rho$ that can be written in a form of a series
\begin{equation}
\hat f(\rho)=\sum_k s_k\rho^k
\end{equation}
transforms in the same way. It follows that all expressions like
\begin{equation}
{\rm Tr}\,\bigl( \hat f(\rho){\bf 1}_I\otimes \hat A_{II}\bigr)
\end{equation}
are $U_I$-independent. Also all functionals depending on reduced
density matrices of subsystems are independent of changes of
bases outside of those subsystems. In the second part of the
paper I shall
prove a general theorem stating that if two observables depend
on reduced density matrices of different subsystems then they
are in involution with respect to a large class of (Poisson or
generalized Nambu) brackets. In
such a case both kinds of telegraphs will be eliminated.

The last point that has to be explained is the question of
uniqueness of the dynamics of subsystems. We have shown
already that a knowledge of observables on pure states of
subsystems does not determine their form if the subsystems are
correlated with something else. Here we will show that
observables that differ only on mixed states will, in general,
generate different evolutions of subsystems. In the
``atom+field" case the evolution of the atomic inversion will
depend on our (arbitrary) choice of the description.

Consider our ``canonical" example (\ref{IandII}) but rewritten
in a form including density matrices. For the sake of clarity
let us also  use the more general matrix
$
\hat \epsilon=
\left(\begin{array}{cc}
\epsilon_1&0\\
0&\epsilon_2
\end{array}\right)
$
istead of $\sqrt{\epsilon}\sigma_3$.

We have now an infinite number of possibilities. The simplest
nontrivial ones are
\begin{equation}
H_1(\rho_1)=E_1 {\rm Tr}\,\rho_1 ,
\qquad
H_2(\rho_2)=E_2 {\rm Tr}\, \rho_2 +\frac{({\rm Tr}\,\rho_2\hat
\epsilon)^2}{{\rm Tr}\,\rho_2} \label{IandII'}
\end{equation}
and
\begin{equation}
H_1(\rho_1)=E_1 {\rm Tr}\,\rho_1 ,
\qquad
H_2(\rho_2)=E_2 {\rm Tr}\, \rho_2 +\frac{({\rm Tr}\,\rho_2\hat
\epsilon)^2}{{\rm Tr}\,\rho_2}
\frac{{\rm Tr}\,(\rho_2^2)}{({\rm Tr}\,\rho_2)^2}
 \label{IandII''}.
\end{equation}
Let $\rho_{kl}=\sum_m \psi_{mk}\bar \psi_{ml}$ denote the
components of the reduced density matrix $\rho_2$. The
Hamiltonian function of the whole system is
\begin{equation}
H(\rho)=H_1(\rho_1)+H_2(\rho_2).
\end{equation}
The two forms of $H_2$ lead, respectively, to the following two
evolution equations
\begin{equation}
\frac{d}{dt}\rho_{kl}=-2i  \frac{{\rm Tr}\,\rho_2\hat \epsilon}
{{\rm Tr}\,\rho_2}
\rho_{kl}(\epsilon_k-\epsilon_l)
\end{equation}
and
\begin{equation}
\frac{d}{dt}\rho_{kl}=-2i\frac{{\rm Tr}\,\rho_2\hat \epsilon}
{{\rm Tr}\,\rho_2}
\frac{{\rm Tr}\,(\rho_2^2)}{({\rm Tr}\,\rho_2)^2}
\rho_{kl}(\epsilon_k-\epsilon_l)
\end{equation}
where the expressions involving traces are integrals of motion.
The equations are different.

There is only one way out of the above dilemma: The description
must from the outset be given in terms of density matrices.

\section{A Two-Level Atom in Nonlinear QM}
\label{2P}

 The
aim of the analysis below is to clarify some elementary features
of the nonlinear formalOAism involved in calculations of
optical pheneomena in two-level atoms.
  The contents of this section should not be
understood as a complete, unique solution or systematization of
all the questions encountered. Still, I hope that I have managed
to point out some elements essential for correct computations in
practical, experimental situations.

In linear QM a two-level system is mathematically equivalent to
a spin-1/2 nonrelativistic particle. All the examples discussed
in this work were based on a two-dimensional Hilbert space and
the reader may have a feeling that they could be applied equally
well to both spin-1/2 particles and two-level atoms. This kind
of conviction has been shared by all the authors dealing with
theoretical and experimental
aspects of nonlinear QM.  The main result of this section, as we
shall see later, is that, paraphrasing G.~Orwell's words, in
nonlinear QM all two-level systems are two-level but some of
them are more two-level than others.

We shall assume that the putative nonlinearity is of purely
atomic origin. This means that we shall consider the atom as a
nonlinear {\it subsystem\/} of the larger ``atom+field" system
where both the field and interaction Hamiltonian functions are
linear in density matrix. (The papers discussing the problem can
be divided into two groups: Either the authors do not care about
the description of the ``atom+field" composite system or treat
it in the way proposed by Weinberg. We know that none of them
can be correct unless the atom and the field are in a product
state, which is typical for semiclassical treatments. It follows
that no really quantum description of the problem has been given
as yet.) We shall assume also the dipole and rotating wave
approximations. To clarify the role of the latter we shall
briefly treat
both $\Delta m=0$ and $\Delta m=\pm 1$ cases.

We begin with the form of the atomic Hamiltonian function. The
simplest one (at lest from the point of view of simplicity of
calculations) is
\begin{equation}
H_{AT}[\rho]={\rm Tr}\, \rho \hat H_L +
\frac{({\rm Tr}\, \rho\hat \epsilon)^2}{{\rm Tr}\,
\rho}\label{atom}
\end{equation}
where $\hat H_L$ is the linear Hamiltonian of the atom and $\hat
\epsilon$ is an operator commuting with $\hat H_L$. Assuming
that we consider the atom in a pure state
$\rho=|\psi\rangle\langle\psi|$ we find that a general solution
of the resulting nonlinear Schr\"odinger equation is (in this
section we shall use the ordinary units with $\hbar\neq 1$)
\begin{equation}
\psi_k(t)=\psi_k(0)\exp\Bigl[-\frac{i }{\hbar}\Bigl(E_k +
2 \langle\hat \epsilon\rangle\epsilon_k -
\langle\hat \epsilon\rangle^2\Bigr)t\Bigr].
\end{equation}
The averages of $\hat \epsilon$ are integrals of motion and, of
course, depend on {\it all\/} nonvanishing components of
$|\psi\rangle$.  This is an important point.  In the analysis of
a coupling between the atom and an external electromagnetic
field we meet two difficulties. First of all we have to decide
which states will be involved in the absorbtion-emission
process. In linear QM the situation is simple: We take two
stationary states of the noninteracting atom. In nonlinear QM
the atomic nonlinearity may lead to stationary states that are
not orthogonal between one another. Atomic creation and
annihilation operators corresponding to such levels cannot
satisfy ordinary anticommutation relations.

To avoid such complications the interaction term we shall choose
will be defined in the ordinary way, that is, in terms of
creation and annihilation operators corresponding to the levels
of the {\it linear\/} Hamiltonian.

Second, we cannot {\it a priori\/} restrict the atomic Hilbert
space to two dimensions because a $k$-th eigenfrequency depends
on amplitudes of all other components of $|\psi\rangle$ and the
evolution cannot be naturally ``cut into $N$-dimensional
pieces". Of course, we cannot also assume that only these
$\epsilon_k$ are nonvanishing which we are interested in (a
probability that in a given case we will find just those only
nonvanishing ones is 0). Therefore, to make the analysis
 perfectly consistent
we should give up the two level approximation in the
interaction term. We shall not, however, consider such
complications although this approximation will further restrict
the physical validity of the calculations presented below.

Let $b_k,\,b^{\dag}_k$ be the $k$-th level atomic annihilation
and creation operators satisfying the fermionic algebra
$[b_k,b^{\dag}_l]_+=\delta_{kl}$ and $a,\,a^{\dag}$ be the
annihilation and creation operators of a monochromatic photon
field whose frequency is $\omega$. The choice of the
creation-annihilation operators language leads naturally to the
following Hamiltonian function of the whole ``atom+field"
composite system
\begin{eqnarray}
H_{A+F}(\psi,\bar \psi)&=&\langle\psi| \sum_k\hbar\omega_k
b^{\dag}_k b_k+
\hbar\omega a^{\dag}a +\frac{i \hbar q}{2}(b^{\dag}_2 b_1
a-a^{\dag} b^{\dag}_1 b_2)|\psi\rangle\nonumber\\ &\phantom{=}&
+
\frac{\langle\psi| \sum_k\epsilon_k b^{\dag}_k
b_k|\psi\rangle^2}{{\langle\psi|\psi\rangle}}.\label{HA+F}
\end{eqnarray}
The state in the Fock basis is
$|\psi\rangle=\sum_{kn}\psi_{kn}| k\rangle| n\rangle$. The nonlinear
term is therefore equivalent to
\begin{equation}
\frac{\langle\psi|\hat \epsilon\otimes {\bf
1}_F|\psi\rangle^2}{{\langle\psi|\psi\rangle}}=
\frac{({\rm Tr}\, \rho_{AT}\hat \epsilon)^2}{{\rm Tr}\,
\rho_{AT}}
\end{equation}
which seems natural but, as we have seen before, is only {\it
one\/} out of the whole variety of inequivalent possibilities.
Such a description is consistent with the definition of the
atomic Hamiltonian function (\ref{atom}) (in this sense it is
unique) and free from any ``malignant" nonlocalities. On the
other hand it is in a mean-field style; we have remarked in the
section on observables that
\begin{equation}
\frac{{\rm Tr}\, \rho_{AT}\hat \epsilon\rho_{AT}\hat
\epsilon}{{\rm Tr}\,
\rho_{AT}}\label{atom'}
\end{equation}
would look more ``fundamental". Anyway, although the conclusions
drawn on a basis of such an analysis are limited in their
generality, some choice has to be made and calculations with
(\ref{atom'}) would be more complicated.

The Hamiltonian resulting from (\ref{HA+F}) is
\begin{eqnarray}
\hat H_{A+F}&=&
\sum_k\hbar\omega_k b^{\dag}_k b_k+
\hbar\omega a^{\dag}a +\frac{i \hbar q}{2}(b^{\dag}_2 b_1
a-a^{\dag} b^{\dag}_1 b_2)\nonumber\\ &\phantom{=}& +
2\frac{\langle\psi| \sum_k\epsilon_k b^{\dag}_k
b_k|\psi\rangle}{{\langle\psi|\psi\rangle}}
\sum_k\epsilon_k b^{\dag}_k b_k-
\frac{\langle\psi| \sum_k\epsilon_k b^{\dag}_k
b_k|\psi\rangle^2}{{\langle\psi|\psi\rangle}^2}
\end{eqnarray}
and the nonlinear Schr\"odinger equation is
\begin{eqnarray}
i \hbar\dot \psi_{1n} &=& \Bigl\{\hbar \omega_1 +2
\frac{\sum_{lm}\epsilon_l|\psi_{lm}|^2}
{\sum_{lm}|\psi_{lm}|^2}\epsilon_1-
\Bigl(\frac{\sum_{lm}\epsilon_l|
\psi_{lm}|^2}{\sum_{lm}|\psi_{lm}|^2}\Bigr)^2+
n\hbar\omega\Bigr\}\psi_{1n}\nonumber\\ &\phantom{=}& - {i\over
2}\hbar q\sqrt{n}\psi_{2,n-1}\nonumber\\ i \hbar\dot \psi_{2n}
&=& \Bigl\{\hbar \omega_2 +2
\frac{\sum_{lm}\epsilon_l|\psi_{lm}|^2}{\sum_{lm}
|\psi_{lm}|^2}\epsilon_2-
\Bigl(\frac{\sum_{lm}\epsilon_l|\psi_{lm}|^2}
{\sum_{lm}|\psi_{lm}|^2}\Bigr)^2+
n\hbar\omega\Bigr\}\psi_{2n}\nonumber\\ &\phantom{=}& + {i\over
2}\hbar q\sqrt{n+1}\psi_{1,n+1}\nonumber\\ &\vdots&\nonumber\\ i
\hbar\dot \psi_{kn} &=& \Bigl\{\hbar \omega_k +2
\frac{\sum_{lm}\epsilon_l|\psi_{lm}|^2}
{\sum_{lm}|\psi_{lm}|^2}\epsilon_k-
\Bigl(\frac{\sum_{lm}\epsilon_l|\psi_{lm}|^2}
{\sum_{lm}|\psi_{lm}|^2}\Bigr)^2+
n\hbar\omega\Bigr\}\psi_{kn}\nonumber\\ &\phantom{=}& {\rm
for}\,k>2.
\end{eqnarray}
Writing $\psi_{kn}=A_{kn}\exp(-i \alpha_{kn}/\hbar)$ we find
that for $k>2$ $A_{kn}={\rm const}$ for all $n$. Since also
$\|\psi\|$ is time independent (hereafter we put $\|\psi\|=1$)
it follows that for $k>2$ the exponents depend on time also {\it
via\/}
$\langle\psi|\epsilon_{12}|\psi\rangle=\sum_n(\epsilon_1|
\psi_{1n}|^2+\epsilon_2|\psi_{2n}|^2)$
whose explicit form has to be determined. Decomposing
\begin{eqnarray}
\langle\psi|\epsilon_{12}|\psi\rangle&=&
{1\over 2}\langle\psi|(\epsilon_1+\epsilon_2)(b^{\dag}_1
b_1+b^{\dag}_2 b_2)|\psi\rangle+ {1\over
2}\langle\psi|(\epsilon_1-\epsilon_2)(b^{\dag}_1 b_1-b^{\dag}_2
b_2)|\psi\rangle\nonumber\\ &=:&
(\epsilon_1+\epsilon_2)\langle\psi|
R_0|\psi\rangle+(\epsilon_2-\epsilon_1)\langle\psi|
R_3|\psi\rangle,
\end{eqnarray}
where the first expression is an integral of motion, we see that
the problem reduces to calculating $\langle\psi|
R_3|\psi\rangle$ which is one half the atomic inversion.
Denoting $\varepsilon= \hat \epsilon\otimes {\bf 1}_F$, the
total Hamiltonian can be decomposed now into two parts
\begin{eqnarray}
\hat H_1&=&\hbar(\omega_2-\omega_1)R_3+
2(\epsilon_2-\epsilon_1)\langle\psi| \varepsilon|\psi\rangle R_3
+\hbar\omega a^{\dag}a\nonumber\\ &\phantom{=}& +\frac{i \hbar
q}{2}(b^{\dag}_2 b_1 a-a^{\dag} b^{\dag}_1 b_2),\nonumber\\
\hat H_2&=&
\hbar(\omega_1+\omega_2)R_0+
\sum_{k>2}\hbar\omega_k b^{\dag}_k b_k+
2\langle\psi| \varepsilon|\psi\rangle
\bigl((\epsilon_1+\epsilon_2)R_0\nonumber\\
&\phantom{=}&+
\sum_{k>2}\hbar\epsilon_k b^{\dag}_k b_k\bigr)-
\langle\psi| \varepsilon|\psi\rangle^2.
\end{eqnarray}
Operators, $R_3$, $R_1={1\over 2}(b^{\dag}_2 b_1+b^{\dag}_1
b_2)$, $R_2={i\over 2}(-b^{\dag}_2 b_1+b^{\dag}_1 b_2)$, $a$ and
$a^{\dag}$ commute with $\hat H_2$ so that the evolution of the
atomic operators $R_j$ is generated by the following nonlinear
generalization of the Jaynes-Cummings Hamiltonian
\begin{equation}
\hat H_1=\hbar\omega_0R_3+
2\epsilon_0\langle\psi| \varepsilon|\psi\rangle R_3 +\hbar\omega
a^{\dag}a +\frac{i \hbar q}{2}(R_+ a-a^{\dag} R_-),
\end{equation}
where $\omega_0=\omega_2-\omega_1$ and
$\epsilon_0=\epsilon_2-\epsilon_1$.  To explicitly distinguish
between initial conditions and dynamical objects we shall
decompose $\langle\psi| \varepsilon|\psi\rangle$ as follows
\begin{equation}
\langle\psi|
\varepsilon|\psi\rangle=\sum_{k>2,n}\epsilon_k|\psi_{kn}|^2+
(\epsilon_1+\epsilon_2)\langle\psi| R_0|\psi\rangle+
\epsilon_0\langle\psi| R_3|\psi\rangle:=A+\epsilon_0\langle\psi|
R_3|\psi\rangle
\end{equation}
where $A$ is a constant depending on initial conditions.
Denoting further
$
\hbar\omega_0+2\epsilon_0A=\hbar\omega_0'$,
$2\epsilon_0^2= \hbar\epsilon\label{A}
$
we finally obtain
\begin{equation}
\hat H_1=\hbar\omega_0'R_3+
\hbar\epsilon\langle\psi| R_3|\psi\rangle R_3
+\hbar\omega a^{\dag}a +\frac{i \hbar q}{2}(R_+ a-a^{\dag} R_-).
\end{equation}
It is clear now what is the actual meaning of the two-level
approximation in nonlinear QM. The evolution of the atomic
operators $R_j$ is generated by a two-dimensional Hamiltonian in
analogy to the linear case, but the parameters of $\hat H_1$
depend on components of the wave function corresponding to the
levels being outside of the two-dimensional Hilbert space. On
the other hand, the phases of the remaining components depend on
the average of the atomic inversion of the two levels.

(It should be stressed  that one can consider a quantum
system whose Hamiltonian function is a sum of a linear term and
some nonlinearity which involves only spinor components of the
wave function. Formally, such a case would be equivalent to a
composite system whose constituents are a scalar particle that
evolves according to the laws of the linear QM and some {\it
nonlinear\/} two-level system noninteracting with the particle.
We know that various solutions of the nonlinear Schr\"odinger
equation, including the product one, will exist and no
components other than the spinor ones will be involved in the
nonlinear part of the evolution.  Systems with nonlinearities of
this kind would be ``truely two-level" as opposed to the systems
which are two-level in the sense specified above.)

Our next task is to find and solve an equation for the atomic
inversion. The Poisson bracket evolution equation reads
\begin{equation}
i \hbar\frac{d}{dt}\langle\psi| R_3|\psi\rangle=\langle\psi|
[R_3, H_1]|\psi\rangle=
\frac{q}{2}\langle\psi| R_+a+a^{\dag}R_-|\psi\rangle
\end{equation}
so is just like in linear QM \cite{AER}. Following the notation
of
\cite{AER} the second derivative is found equal
\begin{eqnarray}
\frac{d^2}{dt^2}\langle\psi| R_3|\psi\rangle &=&
 -q^2
\langle\psi| R_3\bigl(\hat N+{1\over
2}\bigr)|\psi\rangle\nonumber\\
&\phantom{=}&+\frac{i q}{2}
\bigl(\Delta' +\epsilon\langle\psi|
R_3|\psi\rangle\bigr)\langle\psi|
R_+a-a^{\dag}R_-|\psi\rangle\label{2pR_3}
\end{eqnarray}
where $\hat N=R_3+a^{\dag}a$ and $\Delta'=\omega_0'-\omega$.
Define
\begin{equation}
B=\sum_{k>2,n}\hbar\omega_k|\psi_{kn}|^2+
\hbar(\omega_1+\omega_2)\langle\psi| R_0|\psi\rangle.
\end{equation}
$\langle\psi|\hat N|\psi\rangle$, like in the linear case, is
constant. In order to get rid of the average in the last row of
(\ref{2pR_3}) we rewrite the whole Hamiltonian function as
follows
\begin{eqnarray}
H_{A+F}&=&\hbar\omega_{A+F}= A^2 + B +
\hbar\omega\langle\psi|\hat N|\psi\rangle\nonumber\\
&\phantom{=}& + \hbar\Delta'\langle\psi| R_3|\psi\rangle
+\epsilon_0^2\langle\psi| R_3|\psi\rangle^2+
\frac{i \hbar q}{2}\langle\psi| R_+a-a^{\dag}R_-|\psi\rangle.
\end{eqnarray}
Denoting $w=2\langle\psi| R_3|\psi\rangle$,
$\omega_A=A^2/\hbar$, $\omega_B=B/\hbar$,
$\omega_{RWA}=\omega_{A+F}-\omega_B$ we finally get
\begin{eqnarray}
\ddot{w}&=&2\Delta'(\omega_{RWA}-\omega\langle\hat
N\rangle-\omega_A)\nonumber\\
&\phantom{=}&+\bigl(\epsilon(\omega_{RWA}-\omega\langle\hat
N\rangle-\omega_A)-\Delta'\bigr)w - q^2\langle\psi|
R_3\bigl(\hat N +{1\over 2}\bigr)|\psi\rangle
\nonumber\\
&\phantom{=}& - \frac{3}{4}\epsilon\Delta' w^2 -
\frac{\epsilon^2}{8}w^3.
\end{eqnarray}
We have met here the characteristic inconvenience of the Poisson
bracket formalism of nonlinear QM: The nonexistence of the
Heisenberg picture. In the linear case we can solve the
Jaynes-Cummings problem completely, independently of any
particular initial conditions for states. Here the term
$\langle\psi| R_3\bigr(\hat N +{1\over 2}\bigr)|\psi\rangle$
involves correlations between the atom and the field and I have
not managed to express it solely in terms of constants and $w$
unless the state is an eigenstate of $\hat N$, or a
semiclassical decorrelation is assumed. So let initially the
state of the system be a common eigenstate of $R_3$ and
$a^{\dag}a$ with respective eigenvalues $n'$ and $n$.  The
atomic inversion satisfies then the general elliptic equation
\cite{Davies}
\begin{equation}
\ddot{w}=2\Delta'(\Delta'n'+\frac{1}{8}\epsilon)
+\Bigl(\epsilon(\Delta'n'+\frac{1}{8}\epsilon)-\Delta'-
\frac{q^2}{2}\bigl( N +{1\over 2}\bigr) \Bigr)w
- \frac{3}{4}\epsilon\Delta' w^2 -
\frac{\epsilon^2}{8}w^3.\label{ddotw}
\end{equation}

Although we could try to find a general expression for the
atomic inversion following from (\ref{ddotw}) it seems more
instructive to make here some simplifying assumptions. First of
all we can take the ``two-level initial conditions", that is,
assume that the initial state of the system is such that the
only nonvanishing components of the wave function are these with
$k=1,2$. Then $\hbar\Delta'=\hbar\Delta +\epsilon_2^2
-\epsilon_1^2$ where $\Delta=\omega_0-\omega$.  (Let me remark
here that in most of the papers dealing with two-level systems
(cf. \cite{WScully,W2}) their authors assumed that for the
``simplest" nonlinearities $\epsilon_2=0$ which seemed to
suggest that the nonlinearity must shift the resonant frequency.
As we can see, the more symmetric ``$\sigma_3$" choice does not
change the detuning.) Further, choosing the ``detuning"
$\Delta'=0$ and denoting $\epsilon^2/8=2\varsigma^2$ we get
\begin{equation}
\ddot{w}=(2\varsigma^2-\Omega^2)w-2\varsigma^2 w^3.
\end{equation}
This equation can be solved immediately. For example, with the
initial condition $w(0)=-1$ we find
\begin{equation}
w(t)=
\left\{
\begin{array}{ll}
- {\rm cn} (\Omega t,\varsigma/\Omega)& {\rm
for}\quad\Omega>\varsigma\\ - {\rm sech} (\Omega t)& {\rm
for}\quad\Omega=\varsigma\\ - {\rm dn} (\varsigma
t,\Omega/\varsigma)& {\rm for}\quad\Omega<\varsigma
\end{array}\right.
\end{equation}
The result is analogous to this of W\'{o}dkiewicz and Scully
\cite{WScully} who
chose the Bloch equations approach.

It is an appropriate point for a brief comparison of our results
with those of Weinberg who chose his own, basis dependent
description of the ``atom+field" system. Consider the same form
of the atomic nonlinearity. The Hamiltonian function of the
composite system in the {\it Fock basis\/} (this basis was
chosen by Weinberg) is
\begin{eqnarray}
H_{A+F}(\psi,\bar \psi)&=&\langle\psi| \sum_k\hbar\omega_k
b^{\dag}_k b_k+
\hbar\omega a^{\dag}a +\frac{i \hbar q}{2}(b^{\dag}_2 b_1
a-a^{\dag} b^{\dag}_1 b_2)|\psi\rangle\nonumber\\ &\phantom{=}&
+\sum_n
\frac{\langle\psi| \sum_k\epsilon_k b^{\dag}_k b_k \hat
P_n|\psi\rangle^2}
{\langle\psi|\hat P_n|\psi\rangle}.\label{HA+F'}
\end{eqnarray}
where $\hat P_n$ project on $n$-photon states. Assume now that
initially the state of the whole system has only one
nonvanishing component $\psi_{11}$ (one photon and the atom in
the ground state). It follows that only $\psi_{11}$ and
$\psi_{20}$ will appear in the nonlinear Schr\"odinger equation.
Our ``nonmalignant" choice leads to the Schr\"odinger equation
of the form
\begin{eqnarray}
i \hbar\dot \psi_{11} &=& \Bigl\{\hbar \omega_1 +2
\frac{\epsilon_1|
\psi_{11}|^2+\epsilon_2|\psi_{20}|^2}{|\psi_{11}|^2+
|\psi_{20}|^2}\epsilon_1 -
\Bigl(\frac{\epsilon_1|\psi_{11}|^2+
\epsilon_2|\psi_{20}|^2}{|\psi_{11}|^2+
|\psi_{20}|^2}\Bigr)^2+
\hbar\omega\Bigr\}\psi_{11}\nonumber\\
&\phantom{=}& - {i\over 2}\hbar q\psi_{20}\nonumber\\ i
\hbar\dot \psi_{20} &=& \Bigl\{\hbar \omega_2 +2
\frac{\epsilon_1|\psi_{11}|^2+
\epsilon_2|\psi_{20}|^2}{|\psi_{11}|^2+
|\psi_{20}|^2}\epsilon_2-
\Bigl(\frac{\epsilon_1|\psi_{11}|^2+
\epsilon_2|\psi_{20}|^2}{|\psi_{11}|^2+
|\psi_{20}|^2}\Bigr)^2\Bigr\}\psi_{20}
\nonumber\\
&\phantom{=}& + {i\over 2}\hbar q\psi_{11},
\end{eqnarray}
while the Weinberg's one leads to
\begin{eqnarray}
i \hbar\dot \psi_{11} &=& \Bigl\{\hbar \omega_1 +2
\frac{\epsilon_1|\psi_{11}|^2}{|\psi_{11}|^2}\epsilon_1
-
\Bigl(\frac{\epsilon_1|\psi_{11}|^2}{|\psi_{11}|^2}\Bigr)^2+
\hbar\omega\Bigr\}\psi_{11}\nonumber\\
&\phantom{=}& - {i\over 2}\hbar q\psi_{20}\nonumber\\ i
\hbar\dot \psi_{20} &=& \Bigl\{\hbar \omega_2 +2
\frac{\epsilon_2|\psi_{20}|^2}{|\psi_{20}|^2}\epsilon_2-
\Bigl(\frac{\epsilon_2|\psi_{20}|^2}
{|\psi_{20}|^2}\Bigr)^2\Bigr\}\psi_{20}
\nonumber\\
&\phantom{=}& + {i\over 2}\hbar q\psi_{11}.
\end{eqnarray}
Simplifying the fractions we obtain
\begin{eqnarray}
i \hbar\dot \psi_{11} &=& (\hbar \omega_1
+\epsilon_1^2)\psi_{11} - {i\over 2}\hbar q\psi_{20}\nonumber\\
i \hbar\dot \psi_{20} &=& (\hbar \omega_2
+\epsilon_2^2)\psi_{20} + {i\over 2}\hbar q\psi_{11}
\end{eqnarray}
which are {\it linear\/} and the only modification with respect
to ordinary QM is that the energy levels of the atom are the
{\it nonlinear eigenvalues\/} corresponding to the atom in the
absence of radiation. In addition, the ``$\sigma_3"$
nonlinearity satisfies $\epsilon_1^2=\epsilon_2^2$ so that for
this choice of the ``simplest" nonlinearity neither the energy
difference nor the shape of the inversion's oscillation would be
affected, whereas we know already that the ``correct"
Polchinski's description
leads to elliptic oscillations even for $w(0)=-1$ and $N+{1\over
2}=1$.

This result explains the difference between the calculations of
Weinberg and those of W\'odkiewicz and Scully. Both of them are
based, more or less implicitly, on different assumptions about
the description
of composite systems. Nevertheless, it is clear that had we chosen
some other ``correct" form of the total Hamiltonian function, we
would have obtained some other solution for the atomic
inversion.

In the calculations in this section we have not needed any
assumption about the ``smallness" of the nonlinearity. Moreover,
as we have shown before, it is not evident what should be actually
meant by a small nonlinearity. This lack of uniqueness is
related to the fact that there exist singular nonlinearities
that are negligible in the lack of correlations, but can become
dominant if the nonlinear system in question correlates with
something else. Anyway, even ignoring these subtle points it
seems reasonable to expect that a physical nonlinearity, if any,
should be in ordinary situations small in some sense. Therefore,
it becomes interesting to understand in what respect the
solutions we have found depend on approximations. In particular
the role of the rotating wave approximation should be clarified.

The easiest way of doing that is to consider transitions with
the selection rules $\Delta m=\pm 1$ involving circularly
polarized light. It can be shown easily that the only difference
with respect to the $\Delta m=0$ transitions discussed above is
the necessity of substituting $q\bar q$ for $q^2$ in the
equations for $w$, so that no qualitative change in the time
dependence of $w$ will appear.

\section{Wigner's friend revisited}

One of the essential elements of the argumentation against the
description of composite systems proposed by Weinberg was its
dependence on a particular basis in a Hilbert space. In ordinary
quantum situations such a dependence is difficult to accept.
However, the situation changes if we enter the domain of the
measurement theory. Notice, that if the role of the composite
system is played by an
``object" and an ``observer", and the nonlinear evolution occurs in the
state space of the {\em observer\/}, then there exists a
priviliged basis in this space, the {\em pointer basis\/}
\cite{Zurek}, and this is the basis that has to be chosen for
calculation of averages of observed quantities, unless one uses
the many-worlds interpretation, which does not select any basis.

The idea that the domain of (coscious) observations is a natural
arena for a nonlinear evolution of quantum states was introduced
by Wigner in his paradox of a friend\cite{MB}. Wigner considered
a physical system that can be described as ``(observed object+
observer$_1$)+observer$_2$".
The hypothetical nonlinearity was to make the
off-diagonal elements of the density matrix of the
``object+observer$_1$" system disappear, since once the result is
consciously observed, no further interferences should be possible.
Such a goal can be acheived either by a ``collapse" of a state
vector, or in some ``no-collapse" way (decoherence approach).
The sugestion of Wigner, contrary to the oppinion of Penrose
expressed in \cite{Shadows},
belonged to the latter class of theories.

A modification of the observed (linear) system can
exert an influence on the observer; the Gisin's telegraph
belongs to such a class of phenomena, so may not be very
pathological in the context of measurements. On the other hand,
the telegraph based on the mobility phenomenon would lead to
telekinetic phenomena \cite{Jahn}. Indeed, consider a pre-measurement
that produces an entangled state of a system and an observer.
Let the observed system be a two-level atom and an
electromagnetic field, initially in the state
\begin{equation}
| {1}\rangle| {1}\rangle +| {2}\rangle| {0}\rangle =|
{11}\rangle +| {20}\rangle
\end{equation}
where the notation is analogous to this from
 the previous section. Assume that
a Hilbert space of the observer is also spanned by two states
$| \pm\rangle$. After the pre-measurement the state of the joint
system is
\begin{equation}
| {11}\rangle| +\rangle  +| {20}\rangle | -\rangle.
\end{equation}
If now the observer's state space undergoes a nonlinear
evolution with the ``$\sigma_3"$ nonlinearity like in
(\ref{IandII}) then two possibilities occur. First, if only the
$| \pm\rangle$ start to rotate with the mobility frequency then the
atomic reduced density matrix is like in linear QM. However, if
the nonlinearity violates the orthogonality of the states $|
{1}\rangle| +\rangle:=| {1+}\rangle$ and $| {0}\rangle|
-\rangle:=| {0-}\rangle$ (i.e. when
the observer makes both his own consciousness and the
interacting photons evolve in a nonlinear way) then the atom
starts to ``feel" it. It follows that, at least {\it in
principle\/}, a suitable form of his ``own" nonlinearity can
enable the observer influence in a statistically observable way
 a behavior of a random generator just by
watching it! Notice that there is also some limitation on the
possibility of the influence. For consider a situation where
there is a number of intermediate states between the random
generator and the observer, so that the entangled state takes
the form
\[
| 1\rangle| {+_1}\rangle| {+_2}\rangle\dots| {+_n}\rangle|
{+}\rangle+
| 2\rangle| {-_1}\rangle| {-_2}\rangle\dots| {-_n}\rangle|
{-}\rangle.
\]
The random generator will ``feel" the mobility only provided the
mobility will involve the underbraced states
\[
| 1\rangle\underbrace{| {+_1}\rangle| {+_2}\rangle\dots|
{+_n}\rangle| {+}\rangle}+
| 2\rangle\underbrace{| {-_1}\rangle| {-_2}\rangle\dots|
{-_n}\rangle| {-}\rangle}.
\]
In case the nonlinearity is more localized, say
\[
| 1\rangle| {+_1}\rangle\underbrace{| {+_2}\rangle\dots|
{+_n}\rangle| {+}\rangle}+
| 2\rangle| {-_1}\rangle\underbrace{| {-_2}\rangle\dots|
{-_n}\rangle| {-}\rangle}
\]
then the generator's density matrix does not contain the
``malignant" terms (one cannot influence the generator by
watching it on TV; or, it is much easier to influence with my
thoughts my own finger than someone else's).

The above phenomenon is present only if we assume the kind of
description {\it \`{a} la\/} Weinberg.
One may hope that at
least in the ``correct" description no ways of exerting
observer's influence on external world by ``thoughts" or
``intentions" exist.  The following surprising example is a
slightly modified version of the phenomenon noticed by
J.~Polchinski
\cite{Polchinski}. This is the only paradox that I was not able
to eliminate in the Nambu-like generalization of QM proposed in
\cite{II}.

Consider a process involving four steps.
(1) A spin-1/2 ion enters a Stern-Gerlach device, which couples
to the linear spin $\sigma_3$ component and the beam splits into
two, ``$+1$" and ``$-1$" sub-beams.
(2) The ``$+1$" beam evolves freely; in the path of the ``$-1$"
beam a macroscopic observer (or a random generator) takes one of
two actions: (a) does nothing (say,
with probability $\lambda_1$), or
(b) rotates the spin into the ``1" direction with a magnetic
field coupled to $\sigma_2$ (with probability $\lambda_2$).
(3) The two beams are rejoined and the
ion enters a region of field coupled to the nonlinear observable
\begin{equation}
f\frac{({\rm Tr}\,\rho\sigma_1)^2}{{\rm Tr}\,\rho}.
\end{equation}
(4) The observer again measures the
spin with a Stern-Gerlach device coupled to $\sigma_3$.

The steps 1 and 2 prepare the initial condition for the
nonlinear evolution, and the reduced density matrix of the ion
after step 2 is
\begin{equation}
\rho_0=\lambda_1
\left(\begin{array}{cc}
1/2&0\\ 0&1/2
\end{array}\right)+
\lambda_2
\left(\begin{array}{cc}
3/4&1/4\\ 1/4&1/4
\end{array}\right).
\end{equation}
The ion's nonlinear Hamiltonian for the step 3 is
\begin{equation}
\hat H=2f\frac{{\rm Tr}\, \rho\sigma_1}{{\rm Tr}\, \rho}\sigma_1
\end{equation}
where we have assumed the ``correct" form of the evolution, i.e.
that it is generated by the Hamiltonian function depending only
on the reduced density matrix of the ion. The reduced density
matrix satisfies
\begin{equation}
i \dot \rho=2f\frac{{\rm Tr}\, \rho\sigma_1}{{\rm Tr}\,
\rho}[\sigma_1,\rho]
\end{equation}
whose solution is
\begin{eqnarray}
\rho(t)&=&e^{-2i  f\langle\sigma_1\rangle\sigma_1t}
\rho(0)e^{2i  f\langle\sigma_1\rangle\sigma_1t}\nonumber\\
&=&\frac{\lambda_1}{2} {\bf 1}+
\frac{\lambda_2}{4}\Bigl(2 {\bf 1}+\sigma_1 +\sigma_3
\cos2\lambda_2ft +
\sigma_2 \sin2\lambda_2ft \Bigr).
\end{eqnarray}
In the analogous manner we can calculate the evolution of the
projector $P_\pm={1\over 2}({\bf 1}\pm\sigma_3)$. We find in the
``Heisenberg picture" (i.e. we solve the Heisenberg equations of
motion with the nonlinear Hamiltonian)
\begin{equation}
P_\pm(t)={1\over 2}\Bigl({\bf 1} \pm \sigma_3\cos 2\lambda_2ft
\mp \sigma_3\sin 2\lambda_2ft\Bigr).
\end{equation}
The linear case, where in step three the ion couples
with the same coupling constant
linearly to
$\sigma_1$, would yield
\begin{equation}
P_\pm(t)={1\over 2}\Bigl({\bf 1} \pm \sigma_3\cos 2ft \mp
\sigma_3\sin 2ft\Bigr).
\end{equation}
Assuming that the time of the interaction during the third step
satisfies $2ft=\pi$ we get, in the linear case,
\begin{equation}
P_\pm(t)={1\over 2}\Bigl({\bf 1} \mp \sigma_3\Bigr)=P_\mp(0),
\end{equation}
which means that the spin changes its sign during the evolution.
In the nonlinear case, however,
\begin{equation}
P_\pm(t)={1\over 2}\Bigl({\bf 1} \pm \sigma_3\cos \lambda_2\pi
\mp \sigma_3\sin \lambda_2\pi\Bigr)
\end{equation}
hence, in particular, the evolution of $P_+$ depends on
$\lambda_2$ --- the probability of one of the two actions taken
by the observer or the random generator in case the spin turned
out to be $-1$. The result means that the evolution of the ion
depends on ``intentions" of the observer concerning his possible
actions he would have undertaken had the spin of the ion turned
out to be $-1$ --- even in case the spin is $+1$ and the
observer is passive!
For example, for $\lambda_2=0$, that is
when the observer is decided not to take any actions, the spin
state of the ion would remain unchanged. In the opposite case,
$\lambda_2=1$, the evolution would be like in linear QM.

It seems that the essential point of this argumentation is the
assumption that a single member of a beam of ions is described
by the same density matrix as the whole ensemble. This is
exactly opposite to the reasoning leading to Gisin's telegraph.
And I think this example indicates one of the most fundamental
conceptual, or practical, difficulties of nonlinear QM: The fact
that we do not really know how to treat beams of sigle objects.
Intuitively, weak beams should be ``linear" while strong ones
could be, perhaps, ``nonlinear" in some mean-field sense.  The
second hint for futher generalizations of linear QM is a
possibility, suggested by Wigner, that the {\it only\/} domain
of fundamental nonlinearities could be the consciousness of
observer. Then the density matrix representing the consciousness
could evolve nonlinearly and no decompositions of the observer
into sub-ensembles would make any sense. Still, the Polchinski's
phenomenon can describe something like intuition: A perception
of a single event depends on the overall property of an ensemble
of such events since the whole density matrix is involved.

A question that arises immediately is how can we distinguish
between systems that are ``conscious" (observers) and
``non-conscious". If one wants to make such distinctions
more formal and without introducing the state vector reduction
postulate, one has to use the language of the information
theory. In the second part of this paper I shall discuss a
generalization of QM based on R\'enyi entropies. Such entropies
naturally select a class of systems that can
{\it gain\/} information. It will be shown that the possibility
of gaining information by a quantum system is naturally related
to the nonlinearity of evolution.

\acknowledgments
I am grateful to my advisor prof. Kazimierz Rz\c a\.zewski,
prof. Iwo Bia\l ynicki-Birula, and Nicolas Gisin for their
valuable comments
and various help.

\end{document}